%% file: Elementwise_relaxed_Leverage_NNM.tex
\documentclass[11pt]{article}
\usepackage{hyperref}
\usepackage{algorithm,algorithmic}
\usepackage{amsfonts,amsmath, amssymb, bm}
\input{macros_short}

\input{newcmds}

\usepackage{graphicx}
\usepackage{epstopdf}
\usepackage{float}
\usepackage{caption}
\usepackage{subcaption}
\usepackage{multirow}
\begin{document}
\input{title_abstract}
\input{Intro}
\input{main_result}
\input{experiments}
\input{Proof_main_result}
\input{Proofs}
\input{conclusion}
\clearpage 
\bibliography{sparsification} 
\bibliographystyle{plain}
\end{document}

%% file: macros_short.tex
\def\prob#1{{P\left[{#1}\right]}}

\def\x{{\mathbf x}}
\def\y{{\mathbf y}}

\def\dotfil{\leaders\hbox to 1.5mm{.}\hfill}
\newcounter{rmnum}
\def\RN#1{\setcounter{rmnum}{#1}\uppercase\expandafter{\romannumeral\value{rmnum}}}
\def\rn#1{\setcounter{rmnum}{#1}\expandafter{\romannumeral\value{rmnum}}}

%% file: newcmds.tex
\newcommand{\OpNorm }[1]{\mbox{}\left\|#1\right\|_{op}  }

\newcommand{\FNorm }[1]{\mbox{}\left\|#1\right\|_F  }
\newcommand{\FsNorm }[1]{\mbox{}\|#1\|_F  }
\newcommand{\FNormS}[1]{\mbox{}\left\|#1\right\|_F^2}
\newcommand{\NNorm }[1]{\mbox{}\left\|#1\right\|_*  }

\newcommand{\TNorm}[1]{\mbox{}\left\|#1\right\|_2}
\newcommand{\TsNorm}[1]{\mbox{}\|#1\|_2}
\newcommand{\TNormS}[1]{\mbox{}\left\|#1\right\|_2^2}

\newcommand{\VINorm }[1]{\mbox{}\left\|#1\right\|_{\infty}  }
\newcommand{\VsINorm }[1]{\mbox{}\|#1\|_{\infty}  }

\newcommand{\muNorm }[1]{\mbox{}\left\|#1\right\|_{\mu(\infty)}  }

\newcommand{\muTNorm }[1]{\mbox{}\left\|#1\right\|_{\mu(\infty,2)}  }

\newcommand{\setlinespacing}[1]%
           {\setlength{\baselineskip}{#1 \defbaselineskip}}

\newcommand{\abs }[1]{\left|#1\right|}

\newtheorem{definition}{Definition}
\newtheorem{proposition}{Proposition}
\newtheorem{lemma}{Lemma}
\newtheorem{theorem}{Theorem}

\newenvironment{Proof}{\noindent {\em Proof:}}{\\\hspace*{\fill}\mbox{$\diamond$}}

\newcommand{\mat}[1]{{\ensuremath{\bm{\mathrm{#1}}}}}

\def\log{\hbox{\rm log}}

\def\e{{\mathbf e}}

\def\s{{\mathbf s}}
\def\x{{\mathbf x}}
\def\u{{\mathbf u}}
\def\v{{\mathbf v}}

\def\matI{\mat{I}}

\def\matM{\mat{M}}

\def\matS{\mat{S}}

\def\matU{\mat{U}}
\def\matV{\mat{V}}
\def\matW{\mat{W}}
\def\matX{\mat{X}}
\def\matY{\mat{Y}}
\def\matZ{\mat{Z}}
\def\matSig{\mat{\Sigma}}

%% file: title_abstract.tex
\title{\textit{Relaxed} Leverage Sampling for Low-rank Matrix Completion}
\author{
Abhisek Kundu
\thanks{
Research Scientist, 
Xerox Research Center India, 
Bangalore, 
abhisekkundu@gmail.com
}
}

\date{}
\maketitle

\begin{abstract}
\noindent 
We consider exact recovery of any $m\times n$ matrix of rank $\varrho$ from a small number of observed entries via the nuclear norm minimization in (\ref{eqn:main_problem}). Such matrices have degrees of freedom $(m+n)\varrho - \varrho^2$. We show that any low-rank matrix can be recovered exactly from 
$\Theta\left(((m+n)\varrho - \varrho^2)\log^2(m+n)\right)$ randomly sampled entries, thus matching the lower bound  
on the required number of entries (in degrees of freedom), with an additional factor of $O(\log^2(m+n))$.
To achieve this bound we observe each entry
with probabilities proportional to the sum of corresponding row and column leverage scores, \textit{minus their product} (see (\ref{eqn:main_probability})). 
We show that this relaxation in sampling probabilities (as opposed to sum of leverage scores in \cite{BCSW14}) can give us 
$O(\varrho^2\log^2(m+n))$ improvement 
on the (best known) sample size obtained by \cite{BCSW14} for the problem in (\ref{eqn:main_problem}).
Experiments on real data corroborate the theoretical improvement on sample size.

Further, exact recovery of $(a)$ incoherent matrices (with restricted leverage scores), and $(b)$ matrices with only one of the row or column spaces to be incoherent, can be performed using our \textit{relaxed leverage score sampling}, via (\ref{eqn:main_problem}), 
{without knowing the leverage scores a priori}. In such settings also we can  achieve improvement on sample size.


%
\end{abstract}

%% file: Intro.tex
\section{Introduction}

Suppose we have a data matrix $\matM \in \mathbb{R}^{m \times n}$ with incomplete/missing entries, say, we have information about only a small number elements of $\matM$. The \textit{matrix completion} problem (\cite{CR09}) is to predict those missing entries as accurately as possible based on the observed entries. Such partially-observed data may appear in many application domains. For example, in a user-recommendation system (a.k.a \textit{collaborative filtering}) we have incomplete user ratings for various products, and the goal is to make predictions about a user's preferences for all the products (e.g., the \textit{Netflix problem}). Also, the incomplete data could represent partial distance matrix in a sensor network, or missing pixels in digital images because of occlusion or tracking failures in a video surveillance system (\cite{CT10}).
 
More mathematically, we have information about the entries $\matM_{ij}$, $(i, j) \in \Omega$, where $\Omega \subset [m] \times [n]$ is a sampled subset of all entries, and $[n]$ denotes the list $\{1,...,n\}$.  
The problem is to recover the unknown matrix $\matM$ in a computationally tractable way from as few observed entries as possible.
However, without further assumption on $\matM$ it is impossible to predict the unobserved elements from a limited number of known entries. One popular assumption is that $\matM$ has low-rank, say rank $\varrho$. Such matrices have degrees of freedom $(m+n)\varrho - \varrho^2$, i.e., this many parameters control all other entries. 
This implies, if the number of observed entries $s = \abs{\Omega} < (m+n)\varrho - \varrho^2$, there can be infinitely many matrices of rank at most $\varrho$ with exactly the same entries in $\Omega$; therefore, exact recovery of unobserved entries is impossible. So, in general, we need at least $(m+n)\varrho - \varrho^2$ many observed entries for exact matrix completion. 
%
The matrix $\matM$, with the observed entries, can be interpreted as an element in $mn$-dimensional linear space, with available information about $O((m+n)\varrho - \varrho^2)$ coordinates. 
The set of matrices compatible with the observed entries forms a large affine space. Then, exact matrix completion problem is to specify an efficient algorithm which uniquely picks $\matM$ from this high-dimensional affine space (\cite{Gross11}).
Since our target matrix $\matM$ is low-rank, a natural optimization problem for finding $\matM$ would be
to find a matrix with minimum rank complying with the observed entries.
However, minimizing rank over an affine space is known to be NP-hard (\cite{Natarajan95}). \cite{CR09} proposed to solve the heuristic optimization in (\ref{eqn:main_problem}) (surrogate for rank minimization,  \cite{Fazel02}) to recover the low-rank matrix $\matM$.
\begin{eqnarray}\label{eqn:main_problem}
\min_{\matX \in \mathbb{R}^{m\times n}}  \NNorm{\matX}
\quad \text{s.t.} \quad \matX_{ij} = \matM_{ij} \quad (i,j)\in \Omega,
\end{eqnarray}
where the nuclear norm $\NNorm{\matX}$ of a matrix $\matX$ is defined as the sum of its singular values,
$
\NNorm{\matX} = \sum_i \sigma_i(\matX).
$
(\ref{eqn:main_problem}) is a convex optimization problem that is efficiently solvable via semi-definite programming. Exact matrix completion thus becomes proving that the nuclear norm restricted to the affine space has a strict and global minima at $\matM$. That is, if $\matM+\matZ \neq \matM$ is a matrix in the affine space in (\ref{eqn:main_problem}), we need to show
$
\NNorm{\matM+\matZ} > \NNorm{\matM}. 
$
\cite{CR09}, \cite{Gross11}, \cite{Recht09}, \cite{CLMW11} developed the sufficient conditions and probabilistic tools to recover $\matM$ as a unique solution to (\ref{eqn:main_problem}).

One natural question is: which elements of $\matM$ should we observe in (\ref{eqn:main_problem}), i.e., how should we construct the sample set $\Omega$?
We want to define some probabilities on the entries of $\matM$. 
%
Most of the existing work focused on the case when $\Omega$ in (\ref{eqn:main_problem}) is constructed by observing the entries of $\matM$ uniformly randomly (\cite{CR09, Gross11, Recht09, CLMW11}). 
However, this data-oblivious sampling scheme has a cost. 
If the matrix is very sparse, it cannot be recovered using uniform sampling of its entries, unless we observe almost all the entries.
This is because by observing only zeros it is impossible to predict non-zeros of a matrix. This suggests that $\matM$ cannot be in the null space of the sampling operator (to be defined later) extracting the values of a subset of the entries. 
Matrices similar to the above example can be characterized by the structure of their singular vectors. The singular vectors are (closely) `aligned' with the standard basis. 
Therefore,
the components of singular vectors should be sufficiently spread to reduce the number of observations needed to recover a low-rank matrix (\cite{Recht09}). Such restrictions on the row and column spaces of a low-rank matrix are called the \textit{incoherence} assumptions (to be defined later).
\cite{Gross11, Recht09} showed that such restricted class of $n \times n$ matrices of rank $\varrho$ can be recovered exactly, with high probability, by observing as small as $O(n\varrho \text{ log}^2n)$ entries sampled uniformly.
Very recently, \cite{BCSW14} proposed non-uniform probabilities proportional to the sum of row and column leverage scores of $\matM$ to observe its entries (\textit{leveraged sampling}). They eliminated the need for those `incoherence' assumptions, and showed that any arbitrary $n \times n$ matrix of rank $\varrho$ can be recovered exactly, with high probability, from as few as $O(n\varrho \text{ log}^2n)$ observed elements. 

Similar to \cite{BCSW14}, we also incorporate the row and column leverage scores of the recovering matrix $\matM$ into our proposed probability of observing an entry. However, we use a relaxed notion of leverage score sampling. Specifically, we propose to observe an entry with probability proportional to the sum of the corresponding row and column leverage scores, \textit{minus their product}. 
Theorem \ref{thm:main} shows that observing entries according to this \textit{relaxed leverage score sampling} in (\ref{eqn:main_probability}), we can recover \textit{any arbitrary} $m \times n$ matrix of rank-$\varrho$ exactly, with high probability, from $\Theta(((m+n)\varrho - \varrho^2)\log^2(m+n))$ observed entries, via 
(\ref{eqn:main_problem}). This bound on the sample size is optimal (up to $\log^2(m+n)$ factor) in the number of degrees of freedom of a rank-$\varrho$ matrix. Also, this can give us  
$O(\varrho^2\log^2(n))$ improvement 
on the sample size in \cite{BCSW14} for $n \times n$ case. 

For an $n \times n$ matrix $\matM$ of rank-$\varrho$ whose column space is incoherent and row space is arbitrarily coherent, \cite{BCSW14} gives a provable sampling scheme (using leveraged sampling) which requires no prior knowledge of the leverage scores of $\matM$. They show that such  $\matM$ can be recovered exactly, with high probability, using sample size $\Theta(n\varrho \text{ log}^2n)$. We can incorporate our relaxed leverage scores in such setting, with no prior knowledge of leverage scores,  to achieve improvement on the sample size obtained by \cite{BCSW14} while recovering $\matM \in \mathbb{R}^{m \times n}$ exactly with high probability. 
Finally, our notion of relaxation in sampling probabilities can also achieve an improvement on the sample size of \cite{BCSW14} even in case of uniform sampling for incoherent matrices. 
Table \ref{table:compare} summarizes some of the recent results in the literature.

\begin{table*}[!t]
\caption
{
Summary of bound on sample size $s$ for exact recovery of matrix $\matM \in \mathbb{R}^{m \times n}$ of rank $\varrho$
}
\centering
    \begin{tabular}
{| c | c | l | c |}
   \hline
matrix type & probabilities & \qquad \qquad \qquad \qquad 
\qquad bound on $s$ 
& citation \\
\hline \hline
incoherent & uniform & $s \geq O(\tau\cdot(m\wedge n)\varrho\log(m\wedge n))$ & \cite{CR09} \\ \hline
incoherent & uniform & $s \geq O(\max\{\mu_1^2,\mu_0\}(\lambda+\varrho^2)\log^2(2n))$ & \cite{Recht09}  \\ \hline
%
%
any & leveraged  & $\mathbb{E}[s] \geq O((\lambda+\varrho^2)\log^2(m+n))$ & \cite{BCSW14} \\ \hline
any & relaxed leverage  & $\mathbb{E}[s] \leq O(\lambda \cdot \log^2(m+n))$ & Theorem \ref{thm:main} \\ \hline
\multicolumn{3}{@{}l}{$\lambda = (m+n-\varrho)\varrho$,  
$m\wedge n = \max\{m,n\}$, $\tau = \max\{\mu_1^2,\mu_0^{1/2}\mu_1,\mu_0 (m\wedge n)^{1/4}\}$} 
   \end{tabular}
\label{table:compare}
\end{table*}
%
\subsection{Notations and preliminaries}
%
$[n]$ denotes natural number $\{1,...,n\}$. Natural logarithm of $x$ is denoted by $\log(x)$.
Matrices are bold uppercase, vectors are bold lowercase, and scalars are not bold. We denote the $(i,j)$-th entry of a matrix $\matX$ by $\matX_{ij}$. $\e_i$ denotes the $i$-th standard basis vector whose dimension should be clear from the context.
$\matX^T$ and $\x^T$ denote the transpose of matrix $\matX$ and vector $\x$, respectively. $\text{Tr}(\matX)$ denotes the trace of a square matrix $\matX$. 

Spectral norm of $\matX$ is denoted by $\TsNorm{\matX}$. The inner product between two matrices is $\left<\matX,\matY\right> =\text{Tr}(\matX^T\matY)$. Frobenius norm $\matX$ is denoted by $\FNorm{\matX}$, and $\FNorm{\matX} = \sqrt{\left<\matX,\matX\right>}$. 
The maximum entry of $\matX$ is denoted by $\VINorm{\matX} = \max_{i,j}\abs{\matX_{ij}}$. For vectors Euclidean $\ell_2$ norm is denoted by $\TNorm{\x}$.

Linear operators acting on matrices are denoted by calligraphic letters. The spectral norm (largest singular value) of such operator $\mathcal{A}$ will be denoted by $\OpNorm{\mathcal{A}} = \sup_{\matX}\FNorm{\mathcal{A}(\matX)}/\FNorm{\matX}$.
Also, we denote $f(n) = \Theta(g(n))$ when $\alpha_1\cdot g(n) \leq f(n) \leq \alpha_2\cdot g(n)$, for some positive universal constants $\alpha_1, \alpha_2$.

%

%% file: main_result.tex
\section{Main Results}
%

Our focus is to define a probabilities on the entries of $\matM$ (i.e., to construct the sample set $\Omega$ in (\ref{eqn:main_problem})) 
to reduce the sample size, such that $\matM$ becomes the unique optimal solution to (\ref{eqn:main_problem}).
Here we use the Bernoulli sampling model (\cite{CLMW11}), where each entry $(i,j)$ is observed independently with some probability $p_{ij}$. 
Before we state our main result and the distribution, we first need to define the normalized leverage scores (\cite{CR09, Recht09, BCSW14}).\\
\begin{definition}
Let $\matM \in \mathbb{R}^{m \times n}$ be of rank $\varrho$ with SVD $\matM = \matU\matSig\matV^T$, where $\matU$ and $\matV$ are the left and right singular matrices, respectively, and $\matSig$ is the diagonal matrix of singular values. Normalized leverage scores for $i$-th row (denoted by $\mu_i$) and $j$-th column (denoted by $\nu_j$) are defined as follows:
\begin{eqnarray}\label{definition_leverage_scores}
\nonumber
\mu_i &=&  ({m}/{\varrho})\TsNorm{\matU^T\e_i}^2,  \qquad \forall i\in[m], 
\\
\nu_j &=& ({n}/{\varrho})\TsNorm{\matV^T\e_j}^2,  \qquad \forall j\in[n]
\end{eqnarray}
\end{definition}
Normalized leverage scores
\footnote{Leverage scores were introduced by \cite{CH86}$. \TsNorm{\matU^T\e_i}^2$ and $\TsNorm{\matV^T\e_j}^2$ are called row and column leverage scores, respectively, by \cite{DMM08, MD09}}
 are non-negative, and they depend on the structure of row and column spaces of the matrix. Also, we have $\sum_i \frac{\mu_i \varrho}{m} = \sum_j \frac{\nu_j \varrho}{n} = \varrho$, because $\matU$ and $\matV$ have orthonormal columns. 
We state our main result.
\begin{theorem}\label{thm:main}
Let $\matM \in \mathbb{R}^{m \times n}$ of rank $\varrho$. Suppose, we have a subset of observed entries $\Omega \subset [m] \times [n]$, where each entry $(i,j)$ is observed independently with probability $p_{ij}$, such that,
\begin{eqnarray}\label{eqn:main_probability}
%
p_{ij} = \max\left\{\min\left\{c_1 L_{ij}\log^2(m+n), \text{ } 1 \right\}, {(mn)^{-5}}\right\}
\end{eqnarray}
where $L_{ij} = \frac{\mu_i\varrho}{m}+\frac{\nu_j\varrho}{n}-\frac{\mu_i\varrho}{m}\cdot \frac{\nu_j\varrho}{n}$ and $c_1>0$ is some universal constant. Then, $\matM$ is the unique optimal solution to (\ref{eqn:main_problem}) with probability at least $1 - 33\text{ }\log(m+n)(m+n)^{3-c}$, for sufficiently large $c>3$.
Moreover, if the number of observed entries, according to (\ref{eqn:main_probability}), is 
$\Theta\left( ((m+n)\varrho - \varrho^2)\log^2(m+n)\right),$
then, $\matM$ is the unique optimal solution to (\ref{eqn:main_problem}) with probability at least $1 - 66\text{ }\log(m+n)(m+n)^{3-c}$, for sufficiently large $c>3$.
\end{theorem}
Row and column leverage scores measure the contribution of a row or column to the low-rank subspace (\cite{DMM08, MD09}). 
Probabilities in (\ref{eqn:main_probability}) are \textit{biased} towards the leverage score structure of the recovering matrix. This suggests that the elements in important rows and columns, indicated by high leverage scores $\{\mu_i\}$ and $\{\nu_j\}$, of a matrix should be observed more frequently in order to reduce the number of observations needed for exact matrix completion. \cite{BCSW14} also noticed this, and they proposed to sum up $\frac{\mu_i\varrho}{m}$ and $\frac{\nu_j\varrho}{n}$ in the sampling probabilities.
However, our probabilities in (\ref{eqn:main_probability}) reduce this \text{bias} by subtracting the term $\frac{\mu_i\varrho}{m}\cdot \frac{\nu_j\varrho}{n}$ while maintaining the leverage score pattern in $p_{ij}$. This relaxation in probabilities can help us to reduce the number of observations comparing to \cite{BCSW14}, in additive sense, to recover the low-rank matrix exactly, via (\ref{eqn:main_problem}). 

One simple intuition behind this relaxation comes from basic set theory. Let  $\{u_i\}$ and $\{v_j\}$ be the indicators of row and column leverage scores  of the recovering matrix, respectively, where the probabilities are $p(u_i) = \frac{\mu_i\varrho}{m}$ and $p(v_j) = \frac{\nu_j\varrho}{n}$. Then, we want to sample indices $(i,j)$ according to row or column leverage scores, i.e., sampling probabilities $p_{ij}$ to be proportional to $p(u_i \vee v_j) = p(u_i) + p(v_j) - p(u_i \wedge v_j)$. Now, $u_i$ and $v_j$ are independent quantities (an element with high row leverage may or may not have high column leverage, and vice versa). Thus, $p(u_i \wedge v_j) = p(u_i)\cdot p(v_j)$ and $p_{ij} \propto L_{ij}$ in (\ref{eqn:main_probability}).

A practical implication of such relaxation could be as follows. $p_{ij} \propto L_{ij} = \frac{\nu_j\varrho}{n} + (1 - \frac{\nu_j\varrho}{n})\frac{\mu_i\varrho}{m}$. Note that, $0 \leq \frac{\mu_i\varrho}{m}, \frac{\nu_j\varrho}{n} \leq 1$. When $j$-th column is important, i.e.,  $\frac{\nu_j\varrho}{n}$ is high (say close to 1) we want to observe all the elements along that column as $L_{ij} \approx \frac{\nu_j\varrho}{n}$. This eliminates the need for row leverage information of the elements of a column with high leverage score (and vice versa). This reduction of  information for exact completion can lead to a smaller sample set 
in (\ref{eqn:main_problem}).

As discussed earlier, we need a minimum of $\Theta((m+n)\varrho - \varrho^2)$ elements to recover a matrix exactly, regardless of the choice of probabilities. Theorem \ref{thm:main} proves that if we observe elements according to our \textit{relaxed leverage scores}, we match this lower bound, up to a factor of $O(\log^2(m+n))$. 
Here is a comparison of our bound with that of \cite{BCSW14}. 1) We provide a proof for general $m \times n$ matrix of rank-$\varrho$ (as opposed to $n \times n$ case of \cite{BCSW14}),   
2) our form of probabilities establishes an upper bound on the expected sample size 
$\mathbb{E}[s]$.
To see this, note that 
$\sum_{i,j}L_{ij} = (m+n)\varrho-\varrho^2$, and 
$
\mathbb{E}[s] = \sum_{i,j} p_{ij} \leq O(((m+n)\varrho-\varrho^2) \log^2(m+n)),
$
comparing to $\mathbb{E}[s] \geq O(2\max\{m, n\}\varrho\log^2(m+n))$ in \cite{BCSW14}, 
%
3) our sample size to solve (\ref{eqn:main_problem}) can give an improvement on that of \cite{BCSW14} in terms of degrees of freedom (note that $(m+n)\varrho - \varrho^2 < (m+n)\varrho \leq 2\max\{m,n\}\varrho$).


Also, using the relaxed leverage scores we can observe improvement on sample size even in case of \textit{uniform sampling} for matrices with incoherence restrictions. Let $\matM \in \mathbb{R}^{n \times n}$ be the rank-$\varrho$ reconstructing matrix with SVD $\matU\matSig\matV^T$. 
\cite{CR09, CT10, Recht09, Gross11} use two incoherence parameters, $\mu_0$ and $\mu_1$, for exact matrix completion using uniform sampling, 
where, (a) $\max_{i,j}\{\mu_i, \nu_j\} \leq \mu_0$, and (b) $\VsINorm{\matU\matV^T} = \mu_1\sqrt{\varrho/n^2}$. 
A meaningful range of $\mu_0$ is $1 \leq \mu_0 \leq \min\{m,n\}/\varrho$.
\cite{Recht09} showed that if the sampling probability is uniform, such that, 
$
p_{ij} \equiv p \geq c_u{\max\{\mu_0,\mu_1^2\}\varrho \text{ log}^2n}/{n},  \forall i,j,
$
where $c_u$ is a constant, then $\matM$ is the unique optimal solution of (\ref{eqn:main_problem}) with high probability. 
The lower bound achieved on the sample size in \cite{Recht09} (sample-with-replacement model) was $O(\max\{\mu_0,\mu_1^2\}n\varrho \text{ log}^2n)$.
Above, $\mu_1 \leq \mu_0\sqrt{\varrho}$, and it could create a suboptimal dependence of sample size on $\varrho$, in the worst case.
Theorem 2 of \cite{BCSW14} implies that observing entries with uniform probability satisfying,
$
p \geq c_0 \left({2\mu_0\varrho}/{n}\right)\text{log}^2n,  \forall i,j
$,
for some constant $c_0$, would recover the matrix exactly, with high probability. In this case, the lower bound on sample size is $O(2\mu_0n\varrho \text{ log}^2n)$. \cite{BCSW14} eliminated the need for the parameter $\mu_1$, and consequently the suboptimal dependence on $\varrho$.
It follows from Theorem \ref{thm:main} that we can recover the matrix exactly, with high probability, if each entry is sampled uniformly with probability
$$
p = \max\left\{\min\left\{c_1 L_0\text{log}^2(m+n), 1\right\}, (mn)^{-5}\right\},  \forall i,j,
$$
where $L_0 = \frac{\mu_0\varrho}{m} +  \frac{\mu_0\varrho}{n} - \frac{\mu_0^2\varrho^2}{mn}$, $c_1$ is a constant.
This can improve $O(\mu_0^2\varrho^2\text{log}^2(n))$ on the sample size of \cite{BCSW14}.
%

\subsection{Column-Space-Incoherent Matrix Completion}

Here we discuss exact completion of a low-rank matrix whose column space is incoherent, and we have control over the sampling of matrix entries. This setting is interesting in application domains like recommendation systems and gene expression data analysis (\cite{KS13}). 

Algorithm \ref{alg:row_coherent}, adopted from \cite{BCSW14}, performs exact completion of a matrix $\matM$ with incoherent column space, \textit{without a priori knowledge} of leverage scores of $\matM$. 
Step 3 of Algorithm \ref{alg:row_coherent} computes the column leverage scores of $\matM$ exactly, from only a small number of (uniformly) observed rows. 
We construct an additional sample set $\Omega$ of observed entries using our relaxed leverage scores in Step 4. Step 5 solves the nuclear norm minimization problem in (\ref{eqn:main_problem}) with $\Omega$ to recover $\matM$ exactly. Theorem \ref{thm:row_coherent} proves the correctness of Algorithm \ref{alg:row_coherent}.

\input{row_coherent_alg}

\begin{theorem}\label{thm:row_coherent}
Algorithm \ref{alg:row_coherent} computes the column leverage scores of $\matM$ exactly (step 3), i.e., $\tilde{\nu}_j = \nu_j, \forall j \in [n]$. Using the sample set $\Omega$, Algorithm \ref{alg:row_coherent} recovers $\matM$ as the unique optimal solution of (\ref{eqn:main_problem}). The total number of samples required by Algorithm \ref{alg:row_coherent} is 
$
\Theta(\mu_0((m+2n)\varrho - \varrho^2)\log^2(m+n)).
$
The results hold with probability at least, $1 - 66\text{ }\log(m+n)(m+n)^{3-c}$, for sufficiently large $c>3$.
\end{theorem}

We compare the bound on sample size in Theorem \ref{thm:row_coherent} with a couple of existing results. Let us assume $m=n$ for simplicity. Theorem \ref{thm:row_coherent} can achieve an additive improvement $O(\varrho^2\log^2n)$ on the sample size of \cite{BCSW14} while recovering $\matM$ exactly, 
via (\ref{eqn:main_problem}). \cite{KS13} proposed an adaptive sampling algorithm that recovers $\matM$ exactly, with probability at least $1- O(\varrho \delta)$, and a sample size $\Theta(\mu_0 n\varrho^{3/2}\log(\varrho/\delta))$. Assuming comparable failure probabilities, sample size in Theorem \ref{thm:row_coherent} is better when $\varrho$ is not too small.

\subsection{Coherent Matrix Completion using Two-Phase-Sampling}

In reality, we do not have knowledge about the leverage scores of $\matM$, i.e., \{$\mu_i$\} and \{$\nu_j$\}, even when we have control over how to choose entries. \cite{BCSW14} proposed a heuristic two-phase sampling procedure (Algorithm 1 of \cite{BCSW14}) for exact matrix completion with no a priori knowledge about the leverage scores. Here is an informal description of it.

Let, the total budget of samples be $s$, and $\beta \in [0,1]$ be a parameter. First, construct an initial set $\Omega_1$ by sampling entries uniformly (without replacement), such that, $\abs{\Omega_1} = \beta s$. Let $\tilde{\matM}$ be the matrix with $\tilde{\matM}_{ij} = \matM_{ij}$ if $(i,j) \in \Omega_1$, and $\tilde{\matM}_{ij} = 0$ if $(i,j) \notin \Omega_1$.
 Let the rank-$\varrho$ SVD of $\tilde{\matM}$ be $\tilde{\matU}\tilde{\matSig}\tilde{\matV}{}^T$. Compute the leverage scores of $\tilde{\matM}$ and use them as estimates for the leverage scores of $\matM$, i.e., use $\tilde{\mu}_i=\frac{m}{\varrho}\TsNorm{\tilde{\matU}{}^T\e_i}^2$ as $\mu_i$ for $i \in [m]$, and $\tilde{\nu}_j=\frac{n}{\varrho}\TsNorm{\tilde{\matV}{}^T\e_j}^2$ as $\nu_j$ for $j \in [n]$. 
In the second phase, use these estimates to sample (without replacement) remaining $(1-\beta)s$ entries of $\matM$ with probabilities proportional to $(\tilde{\mu}_i\varrho/m+\tilde{\nu}_j\varrho/n) \text{ log}^2(m+n)$, to form the sample set $\Omega_2$. Then perform matrix completion using sample set $\Omega = \Omega_1 \cup \Omega_2$ in (\ref{eqn:main_problem}).

This heuristic is shown to work well on synthetic data that are less coherent (\cite{BCSW14}). For highly coherent data, e.g., only few entries are non-zeros and others are zeros, it works poorly, as expected.
We can incorporate our notion of relaxed leverage scores into the second phase of the above procedure by observing (without replacement) the remaining $(1-\beta)s$ entries of $\matM$ with probabilities 
$
p_{ij} \propto 
(
\frac{\tilde{\mu}_i\varrho}{m} + \frac{\tilde{\nu}_j\varrho}{n} 
- \frac{\tilde{\mu}_i\varrho}{m} \cdot \frac{\tilde{\nu}_j\varrho}{n}
)\log^2(m+n)
$
to form sample set $\hat{\Omega}_2$, and perform nuclear norm minimization in (\ref{eqn:main_problem}) using $\Omega = \Omega_1\cup \hat{\Omega}_2$.
We expect our relaxed leverage score sampling to follow similar trend as above, although we do not evaluate this heuristic numerically.\\

Section \ref{sec:exp} shows experimental results on real datasets to support the theoretical gain on the sample size using relaxed leverage score sampling. 
We give proof sketch of Theorem \ref{thm:main} and Theorem \ref{thm:row_coherent} in Section \ref{sec:thm1} and Section \ref{sec:thm2}, respectively, closely following the proof outline of \cite{BCSW14}. We highlighted the main technical differences between our result and \cite{BCSW14} in Section \ref{sec:thm1}. 

%% file: row_coherent_alg.tex
\begin{algorithm}[!t]
\centerline{
\caption{Column-Space-Incoherent MC}\label{alg:row_coherent}
}
\begin{algorithmic}[1]
\STATE \textbf{Input:} $\matM \in \mathbb{R}^{m \times n}$, with $\max_i \mu_i \leq \mu_0$, $\forall i \in [m]$, 
s.t. $1 \leq \mu_0 \leq m/\varrho$.
\STATE Observe all the entries of a row of $\matM$ picked with probability 
$
p = \min\left\{({c_2\mu_0\varrho\text{ log } m})/{m}, 1\right\},
$
where $c_2$ is a constant.
\STATE Compute the leverage scores, $\{\tilde{\nu}_j\}\text{ } \forall j\in [n]$, of the space spanned by these rows, and use them as estimates for true $\{\nu_j\}, \forall j \in [n]$ of $\matM$. 
\STATE Construct a sample set $\Omega$ of entries $(i,j)$ of $\matM$ observed with probabilities 
\begin{eqnarray}\label{eqn:pij_alg1}
p_{ij} = \min\left\{c_1\tilde{L}_{ij}\log^2(m+n), 1\right\}, \quad \forall i, j,
\end{eqnarray}
where $\tilde L_{ij}=\frac{\mu_0\varrho}{m}+\frac{\tilde{\nu}_j\varrho}{n} - \frac{\mu_0\varrho}{m}\cdot \frac{\tilde{\nu}_j\varrho}{n}$.
\STATE Solve (\ref{eqn:main_problem}) using sample set $\Omega$, and let $\matX^*$ be the unique optimal solution.
\STATE \textbf{Output:} $\matX^*$.
\end{algorithmic}
\end{algorithm} 

%% file: experiments.tex
\section{Experiments}\label{sec:exp}
We show experimental performance of the exact recovery of real data matrices via nuclear norm minimization in (\ref{eqn:main_problem}) using our relaxed leverage score sampling. We use the software `TFOCS' v1.2, written by Stephen Becker, Emmanuel Candes, and Michael Grant, to solve (\ref{eqn:main_problem}).


Let $\matM$ be the rank-$\varrho$ data matrix. We form the sample set $\Omega_{relax}$ by observing $(i,j)$-th entry of $\matM$  according to the relaxed leverage score probabilities in (\ref{eqn:relax_prob_simple}):
\begin{eqnarray}\label{eqn:relax_prob_simple}
p_{ij}^{[relax]} = \min\left\{c_r \cdot L_{ij}, 1\right\},  \forall i, j
\end{eqnarray}
where $L_{ij} = \frac{\mu_i\varrho}{m}+\frac{\nu_j\varrho}{n} - \frac{\mu_i\varrho}{m}\cdot\frac{\nu_j\varrho}{n}$ and $c_r$ is a universal constant.
Similarly, we form the sample set $\Omega_{lev}$ by observing $\matM_{ij}$ according to the leverage score probabilities in (\ref{eqn:lev_prob_simple}): 
\begin{eqnarray}\label{eqn:lev_prob_simple}
p_{ij}^{[lev]} =  \min\left\{c_l \cdot \left(\frac{\mu_i\varrho}{m}+\frac{\nu_j\varrho}{n}\right), 1\right\}, \quad \forall i, j
\end{eqnarray}
where $c_l$ is a universal constant. 
We use $\Omega_{relax}$ and $\Omega_{lev}$ in the optimization problem (\ref{eqn:main_problem}), separately, to recover $\matM$.  Let $\matX^*$ be the optimal solution to (\ref{eqn:main_problem}) using a sample set $\Omega$. We say $\matX^*$ recovers $\matM$ exactly if $\FNorm{\matM - \matX^*} < \varepsilon$, where $\varepsilon$ is a tiny fraction. We set $\varepsilon = 0.001$. We perform 10 independent trials (sampling and recovery) and declare success if $\matM$ is recovered exactly at least 9 times. 
Let $s_r$ and $s_l$ be the average sample size for successful recovery of $\matM$ using $\Omega_{relax}$ and $\Omega_{lev}$, respectively.
We expect $c_r \approx c_l$, and our gain in sample size $(s_l - s_r)$ 
to be strictly positive, as suggested by the theory. Further, we investigate how $(s_l - s_r)$ behaves with respect to the rank $\varrho$.
For this, we define 
\begin{eqnarray}\label{eqn:gain_sample}
\textit{Normalized Gain } (\Delta_s) = \sqrt{{(s_l - s_r)}/{c_r}}.
\end{eqnarray}
We expect ${\Delta_s}$ to be close to $\varrho$ as the theory suggests $(s_l - s_r) \propto O(\varrho^2)$. 
For fairness of comparison, we use the same random seed for both the sampling methods in (\ref{eqn:relax_prob_simple}) and (\ref{eqn:lev_prob_simple}). 
%

\subsection{Datasets}
We use the following two datasets.

\textbf{MovieLens:} \quad
This collaborative filtering dataset contains 100,000 ratings in the range 1 and 5 by 943 users on 1682 movies.
Each user has rated at least 20 movies. This dataset is numerically not low-rank. We perform rank truncation to create an explicit low-rank matrix to apply the theory in (\ref{eqn:main_problem}). We observe the singular value spectrum of this data to heuristically choose two values for rank: $\varrho=10$ and $\varrho=20$.

\textbf{TechTC:}\quad 
We use a dataset from the Technion Repository of Text Categorization Database (TechTC) (\cite{GM04}). Here each row is a document describing a topic, and words (columns) are the features for the topics. The $(i,j)$-th entry of this matrix is the frequency of $j$-th word appearing in $i$-th document. We choose a dataset containing the topics with IDs 11346 ans 22294. We preprocessed the data by removing all words of length four or less. Then, each row is normalized to have unit norm. Also, we observe the singular value spectrum of this preprocessed $125 \times 14392$ data to heuristically choose two values for rank: $\varrho=10$ and $\varrho=20$, to make the data explicitly low-rank. 

\subsection{Results}
\begin{figure}[!h]
\centering
	\includegraphics[height=3.0cm,width=4.0cm]{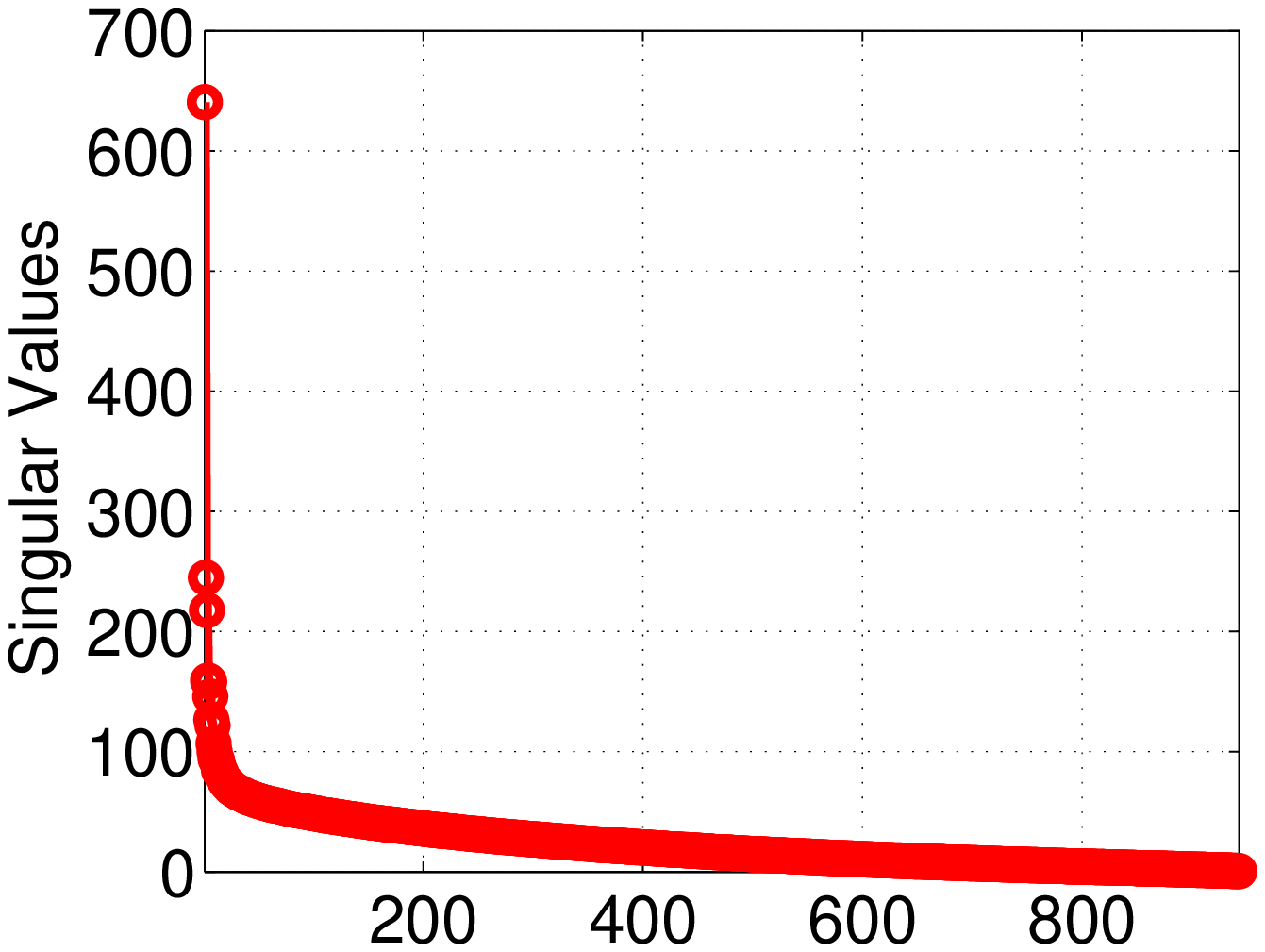}
	\includegraphics[height=3.0cm,width=4.0cm]{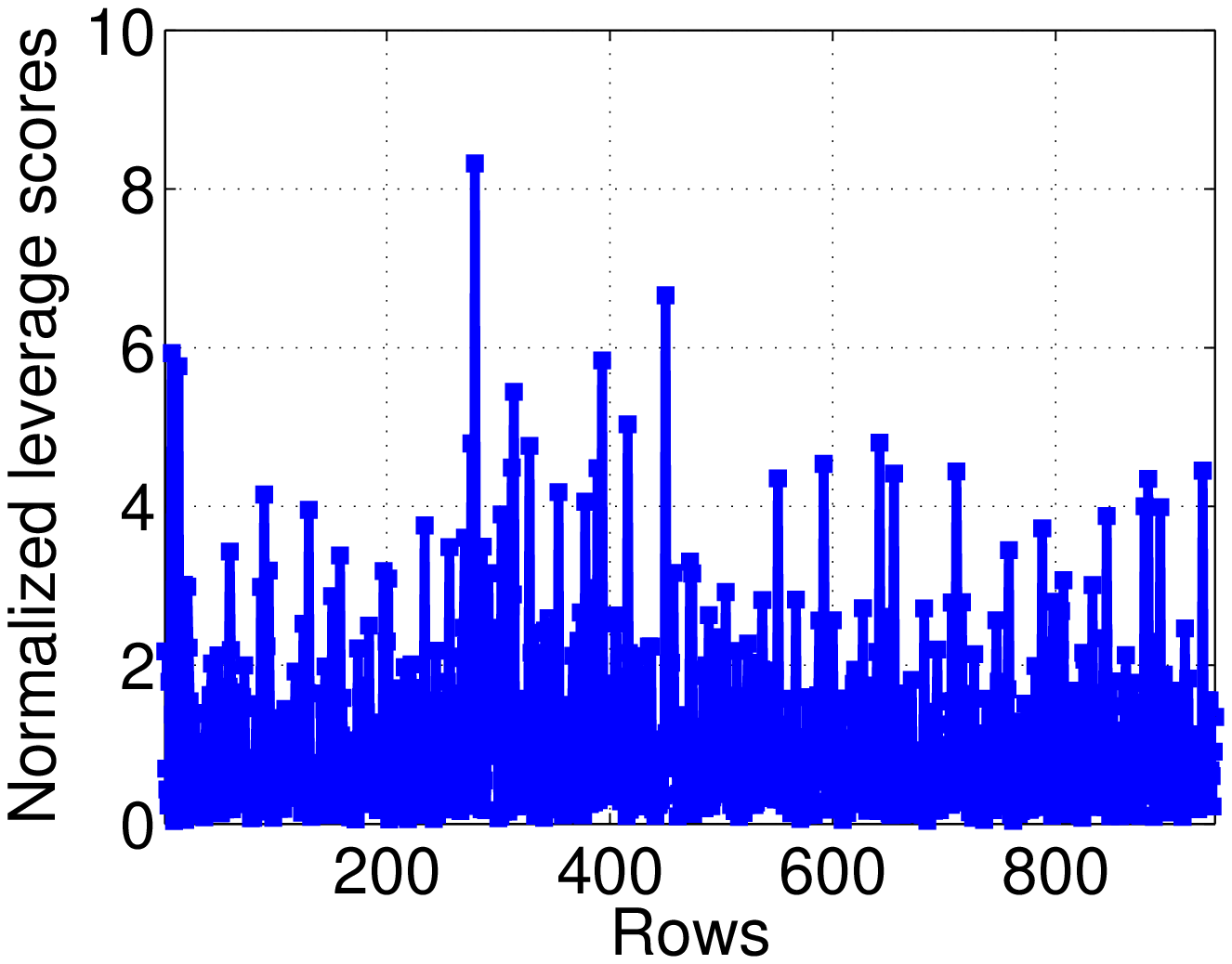}
	\includegraphics[height=3.0cm,width=4.0cm]{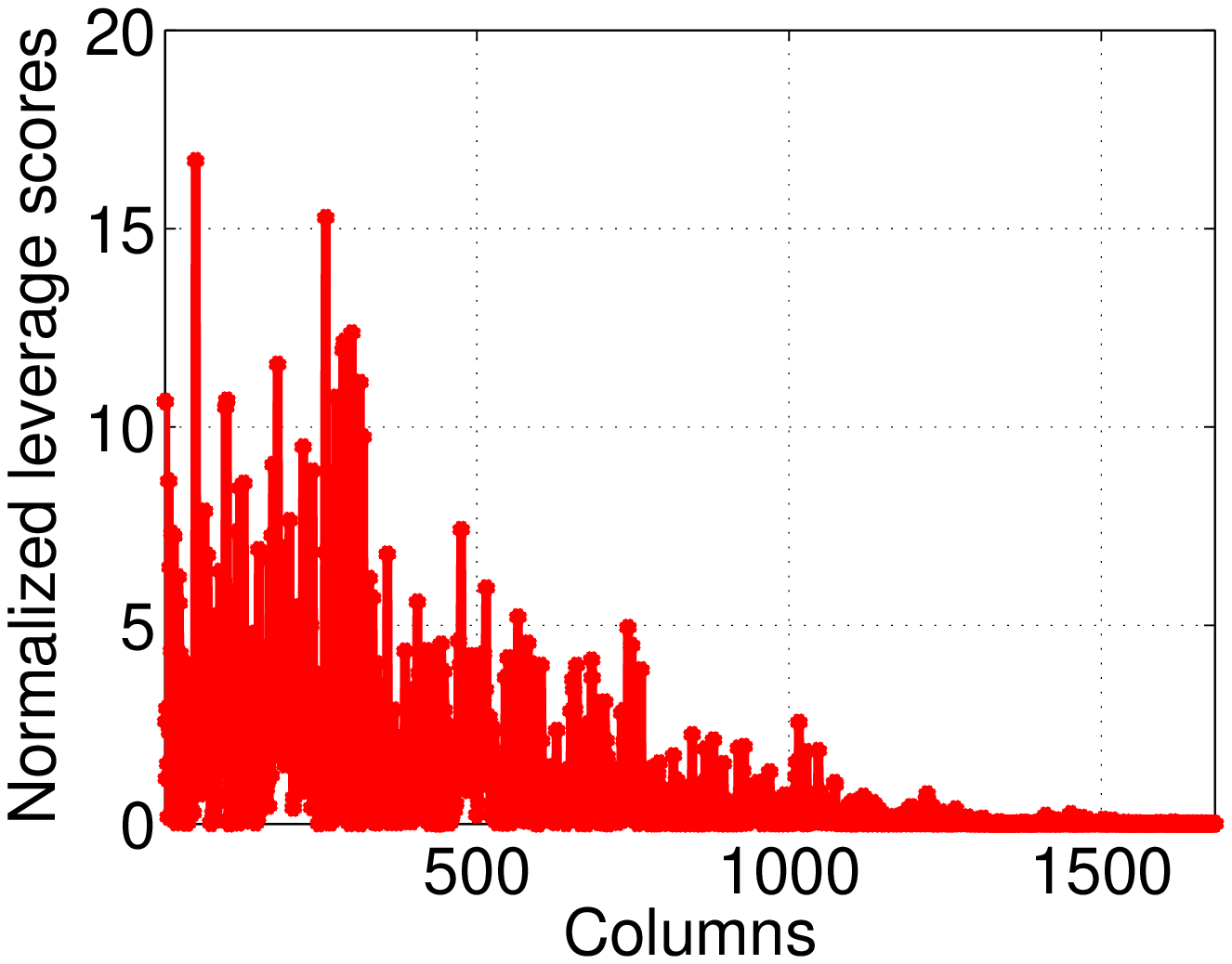}
\caption
{
[MovieLens] Singular values and the leverage scores for $\varrho=10$.
}
\label{fig:movie_leverage}
\end{figure}

\begin{figure}[!h]
\centering
	\includegraphics[height=3.0cm,width=4.0cm]{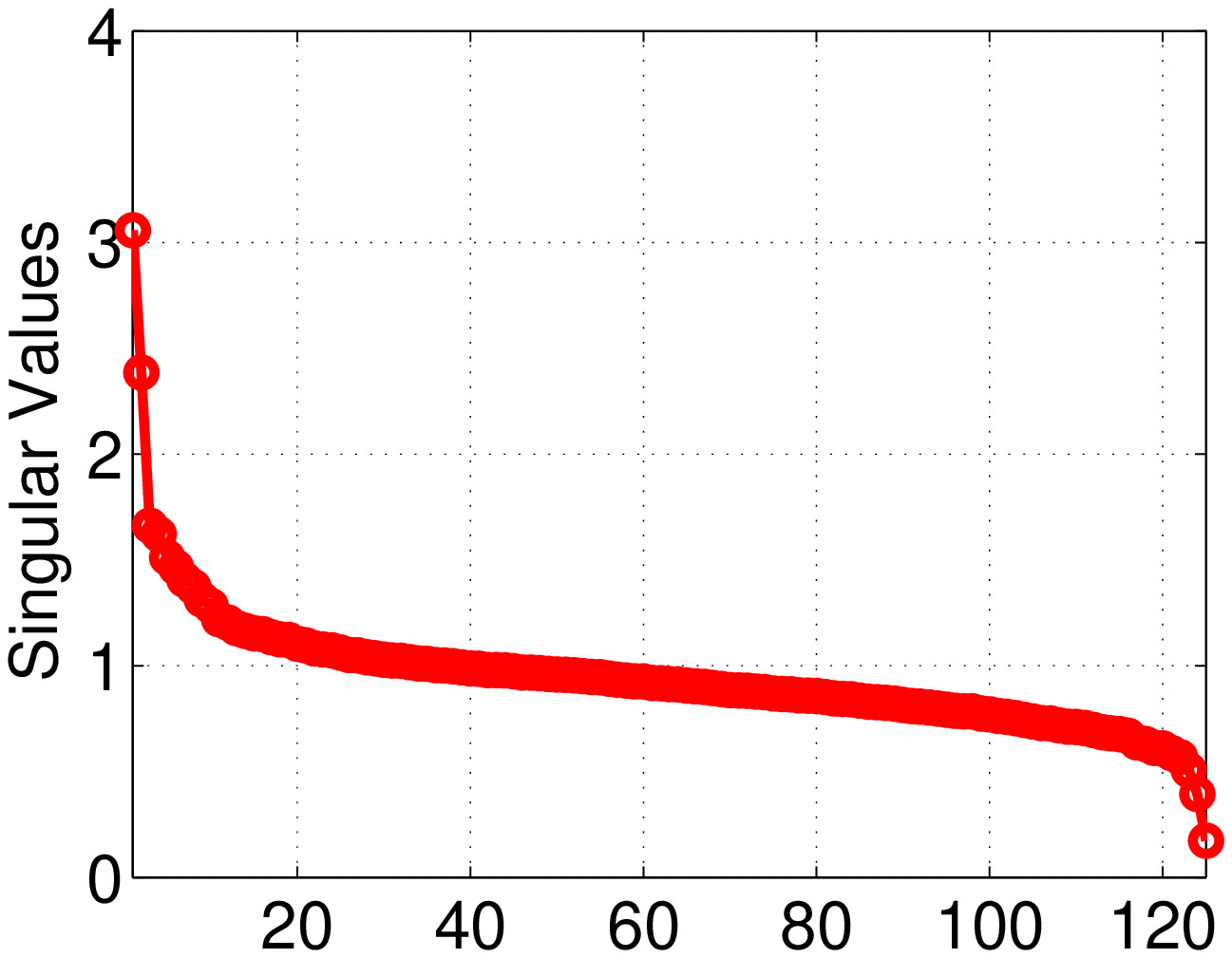}
	\includegraphics[height=3.0cm,width=4.0cm]{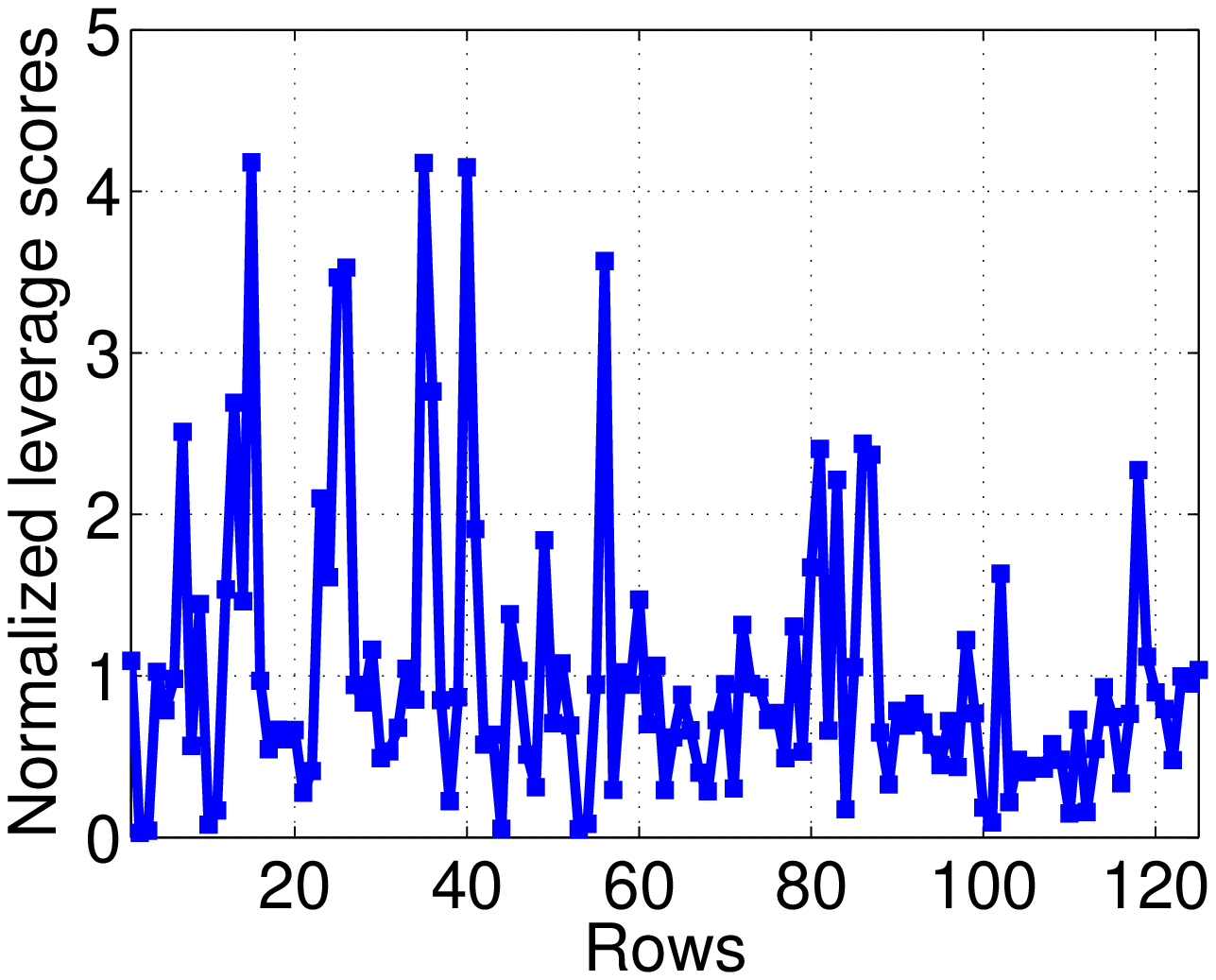}
	\includegraphics[height=3.0cm,width=4.0cm]{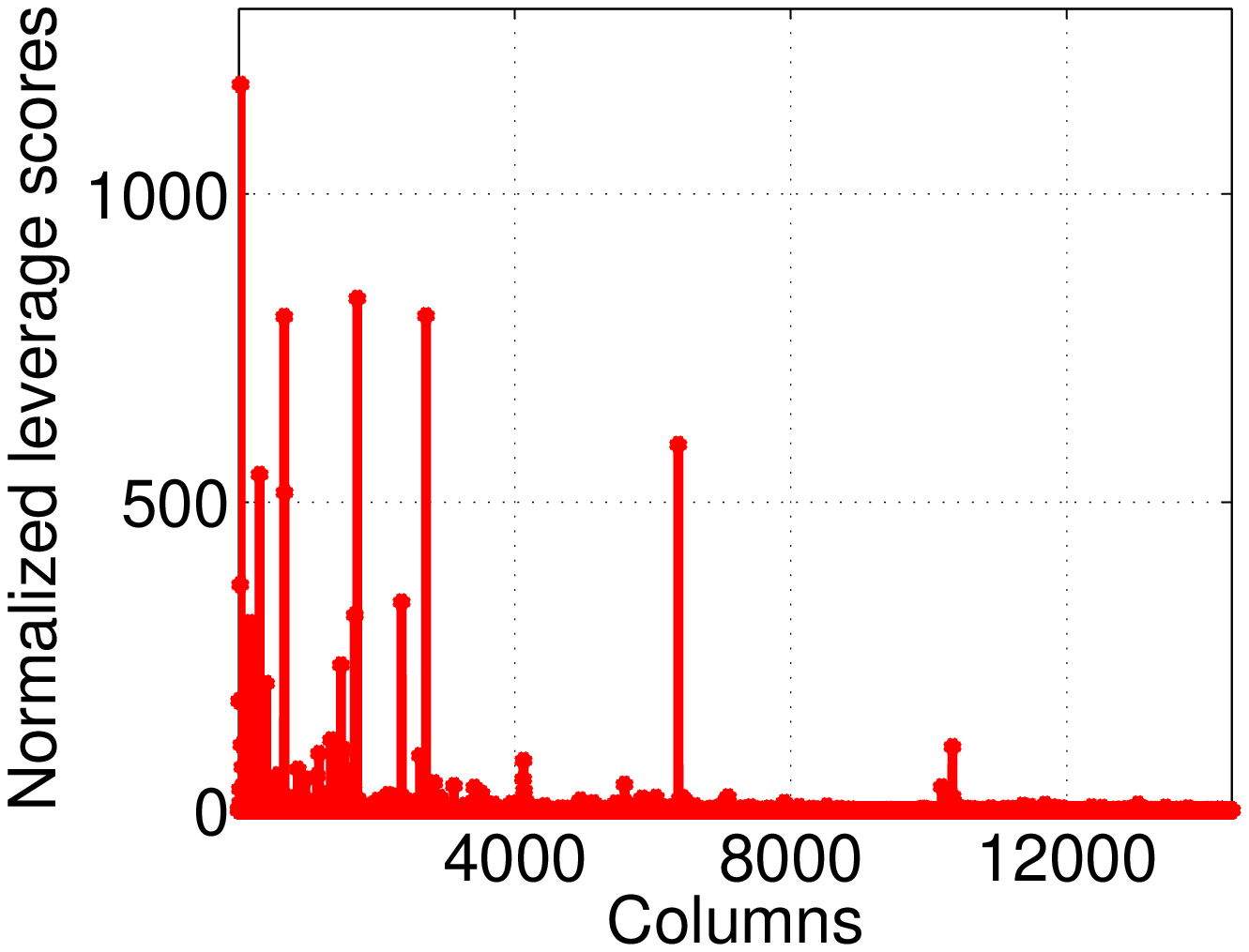}
\caption
{
[TechTC] Singular values, and the leverage scores for $\varrho=10$.
}
\label{fig:techtc_leverage}
\end{figure}

%
Figures \ref{fig:movie_leverage} and \ref{fig:techtc_leverage} plot the singular values and the normalized leverage scores for rank-10 approximation for MovieLens and TechTC data, respectively. Normalized leverage scores are close to 1 when they are incoherent in nature. MovieLens is reasonably coherent, and TechTC has extremely high coherence. 
%
%
%
Table \ref{fig:movie_gain} shows the constants $c_l$ and $c_r$, and the normalized gain $\Delta_s$ for exact recovery of MovieLens data. We see $c_l = c_r$ and $\Delta_s \approx \varrho$, as expected. We observe similar results for TechTC data in  Table \ref{fig:techtc_gain}.
%
\begin{table}[!h]
\begin{minipage}{.45\linewidth}
\centering
    \begin{tabular}{| c | c | r |}
   \hline
& $c_l / c_r$ 
& $\Delta_s$\\
\hline \hline
$\varrho = 10$ & 11/11 &  9.7  \\ \hline
$\varrho = 20$ &  7/7& 18.4 \\ \hline
   \end{tabular}
\caption
{
[MovieLens] Gain in sample size
for exact recovery using relaxed leverage score sampling.
}
\label{fig:movie_gain}
\end{minipage}%
\qquad
\begin{minipage}{.45\linewidth}
\centering
  \begin{tabular}{| c | c | r |}
   \hline
& $c_l / c_r$ 
& $\Delta_s$ \\
\hline \hline
$\varrho = 10$ &  4/4 & 6.6 \\ \hline
$\varrho = 20$ &  3/3 & 15.2\\ \hline 
\end{tabular}
\caption
{
[TechTC] Gain in sample size for exact recovery using relaxed leverage score sampling.
}
\label{fig:techtc_gain}
\end{minipage}%
\end{table}
%
Overall, these results support the accuracy of the theoretical analysis on the gain in sample size using the relaxed leverage score sampling for exact recovery of a low-rank matrix via (\ref{eqn:main_problem}).

%% file: Proof_main_result.tex
\section{Proof of Theorem \ref{thm:main}}\label{sec:thm1}

The main proof strategy was outlined by \cite{CR09, Recht09, Gross11}: it is sufficient to construct a \textit{dual certificate} $\matY$ obeying specific sub-gradient inequalities in order to show that $\matM$ is the unique optimal solution to (\ref{eqn:main_problem}) (see Section \ref{sec:optimality} for more detail). We give a proof of Theorem \ref{thm:main} closely following the proof strategy of \cite{Recht09, BCSW14}. 
Before stating the optimality conditions we need additional notations.

Recall, $\matU$ and $\matV$ are the left and right singular matrices of $\matM$, respectively. Let $\u_k$ (respectively $\v_k$) denote the $k$-th column of $\matU$ (respectively $\matV$). Let $T$ be a linear space spanned by elements of the form $\u_k\y^T$ and $\x\v_k^T$, $1\leq k \leq \varrho$, for arbitrary $\x, \y$, and $T^\perp$ be its orthogonal complement, i.e., $T^\perp$ is spanned by the family $(\x\y^T)$, where $\x$ (respectively $\y$) is any vector orthogonal to the space spanned by the left singular vectors (respectively right singular vectors). Then, orthogonal projection onto $T$ is given by the linear operator $\mathcal{P}_T: \mathbb{R}^{m \times n} \rightarrow \mathbb{R}^{m \times n}$, defined as 
$$
\mathcal{P}_T(\matX) = \matU\matU^T\matX + \matX\matV\matV^T - \matU\matU^T\matX\matV\matV^T.
$$
Similarly, orthogonal projection onto $T^\perp$ is
$$
\mathcal{P}_{T^\perp}(\matX) = \matX - \mathcal{P}_{T}(\matX) = \matU_\perp\matU_\perp^T\matX\matV_\perp\matV_\perp^T.
$$
Note that any $m \times n$ matrix $\matX$ can be expressed as a sum of rank-one matrices as follows:
\begin{eqnarray}\label{eqn:sum_of_terms}
\matX = \sum_{i,j=1}^{m,n}\left<\e_i\e_j^T,\matX \right>\e_i\e_j^T.
\end{eqnarray}
We define the sampling operator $\mathcal{R}_{\Omega}: \mathbb{R}^{m \times n} \rightarrow \mathbb{R}^{m \times n}$ as,
\begin{eqnarray}\label{eqn:sampling_operator}
\mathcal{R}_{\Omega}(\matX) = \sum_{i,j=1}^{m,n}\frac{1}{p_{ij}}\delta_{ij}\left<\e_i\e_j^T,\matX \right>\e_i\e_j^T
\end{eqnarray}
where, $\delta_{ij} = \mathbb{I}((i,j)\in \Omega)$, $\mathbb{I}(\cdot)$ being the indicator function.
That is, $\mathcal{R}_{\Omega}$ extracts the terms, corresponding to the indices $(i,j) \in \Omega$, from (\ref{eqn:sum_of_terms}) to form a partial sum in (\ref{eqn:sampling_operator}).
Let $\mathcal{P}_{\Omega}(\matX)$ be the matrix with $(\mathcal{P}_{\Omega}(\matX))_{ij} = \matX_{ij}$ if $(i,j) \in \Omega$, and zero otherwise.
%
%

\subsection{Optimality Conditions}

Following the proof road map of \cite{Recht09, BCSW14}, we restate the sufficient conditions for $\matM$ to be the unique optimal solution to (\ref{eqn:main_problem}) (Section \ref{sec:optimality} contains a proof of sufficiency).
%
%
\begin{proposition}\label{prop:main}
%
The rank-$\varrho$ matrix $\matM \in \mathbb{R}^{m \times n}$ with SVD $\matM = \matU\matSig\matV^T$ is the unique optimal solution to (\ref{eqn:main_problem}) if the following conditions hold:
\begin{eqnarray*}
1. && \OpNorm{\mathcal{P}_T\mathcal{R}_{\Omega}\mathcal{P}_T - \mathcal{P}_T} \leq {1}/{2}.\\
2. && \text{There exists a dual certificate $\matY$ which satisfies $\mathcal{P}_{\Omega}(\matY) = \matY$, and}
\\
&& (a) \FNorm{\mathcal{P}_T(\matY) - \matU\matV^T} \leq \sqrt{\varrho (m+n)^{-15}},\\
&& (b) \TNorm{\mathcal{P}_{T^\perp}(\matY)} \leq 1/2.
\end{eqnarray*}
\end{proposition}
%
Condition 1 of Proposition \ref{prop:main} suggests $\mathcal{R}_{\Omega}$ should be nearly the identity operator on the subspace $T$. 
Next we discuss the construction of a dual certificate $\matY$.
\subsubsection{Constructing the Dual Certificate}
%
We follow the so-called golfing scheme \cite{Gross11, CLMW11, BCSW14} to construct a matrix $\matY$ (the dual certificate) that satisfies Condition 2 in Proposition \ref{prop:main}. 
Recall, we assume that the set of observed elements $\Omega$ follows the Bernoulli model with parameter $p_{ij}$, i.e., each index $(i,j)$ is observed independently with $\prob{(i,j) \in \Omega} = p_{ij}$ ($p_{ij}$ in eqn (\ref{eqn:main_probability})).
We denote this by $\Omega \sim Bernoulli(p_{ij})$.
Further, we assume that $\Omega$ is generated from  $\Omega = \cup_{k=1}^{k_0}\Omega_k$, where for each $k$,  $\Omega_k \sim Bernoulli(q_{ij})$, and
we set $q_{ij} = 1 - (1-p_{ij})^{1/k_0}$. Clearly, this implies $\prob{(i,j)\in \Omega} = p_{ij}$ which is the original Bernoulli model for $\Omega$. Note that, $q_{ij} \geq p_{ij}/k_0$ because of overlapping of $\Omega_k$'s. 
We set $k_0=11\cdot \log(m+n)$.
Then,
\begin{eqnarray}\label{eqn:probability_qij}
q_{ij} \geq 
\min\left\{c_0\cdot \log(m+n)\cdot \left(\frac{\mu_i\varrho}{m}+\frac{\nu_j\varrho}{n} - \frac{\mu_i\varrho}{m}\cdot \frac{\nu_j\varrho}{n}\right),1\right\},
\end{eqnarray}
where $c_0 = c_1/11$.
Starting with $\matW_0 = 0$ and for each $k=1,..., k_0$, we recursively define 
%
%
\begin{eqnarray}\label{eqn:dual_certificate}
\matW_k = \matW_{k-1} + \mathcal{R}_{\Omega_k}\mathcal{P}_T(\matU\matV^T - \mathcal{P}_T(\matW_{k-1}))
\end{eqnarray}
where the sampling operator $\mathcal{R}_{\Omega_k}: \mathbb{R}^{m \times n} \rightarrow \mathbb{R}^{m \times n}$ is defined as
\begin{eqnarray*}
\mathcal{R}_{\Omega_k}(\matX) = \sum_{i,j}\frac{1}{q_{ij}}\mathbb{I}((i,j)\in \Omega_k)\left<\e_i\e_j^T,\matX \right>\e_i\e_j^T.
\end{eqnarray*}
We set $\matY= \matW_{k_0}$. This $\matY$ is supported on $\Omega$, i.e., $\mathcal{P}_{\Omega}(\matY) = \matY$. 
\\Let the sample set $\tilde{\Omega}$ be such that
\begin{eqnarray}\label{eqn:omega_tilde}
\tilde{\Omega} \in \{\Omega_k:\Omega = \cup_{k=1}^{k_0}\Omega_k, \Omega_k \sim Bernoulli(q_{ij})\}.
\end{eqnarray}
Since $\Omega_k \sim Bernoulli(q_{ij})$ implies $\Omega \sim Bernoulli(p_{ij})$, for each $k=1,...,k_0$, we prove (in Lemma \ref{thm:operator}) Condition 1 of Proposition \ref{prop:main} using sample set $\tilde{\Omega}$ in (\ref{eqn:omega_tilde}).
%
%
\begin{lemma}\label{thm:operator}
Let $\tilde{\Omega}$ be a sample set in (\ref{eqn:omega_tilde}). Then, for any universal constant $c>1$, we have 
%
%
\begin{eqnarray}
\OpNorm{\mathcal{P}_T\mathcal{R}_{\tilde\Omega}\mathcal{P}_T - \mathcal{P}_T} \leq \frac{1}{2}
\end{eqnarray}
holding with probability at least 
$$1 - (m+n)^{1-c}.$$
\end{lemma}
Before we validate Condition 2 in Proposition \ref{prop:main} using the $\matY$ constructed above, we claim the following results to hold with high probability. 
%
First, we borrow the following definitions of weighted infinity norms for a matrix $\matZ \in \mathbb{R}^{m \times n}$ from \cite{BCSW14}. 
\begin{eqnarray*}
\nonumber \muTNorm{\matZ} 
&:=& \max\left\{\max_i\sqrt{\frac{m}{\mu_i \varrho}}\TNorm{\matZ_{i,*}},\max_j\sqrt{\frac{n}{\nu_j \varrho}}\TNorm{\matZ_{*,j}}\right\}\\
%
%
%
\muNorm{\matZ} &:=& \max_{i,j}\abs{\matZ_{ij}}\sqrt{\frac{m}{\mu_i\varrho}}\sqrt{\frac{n}{\nu_j\varrho}}
%
\end{eqnarray*}
where $\matZ_{i,*}$ and $\matZ_{*, j}$ denote the $i$-th row and $j$-th column of $\matZ$, respectively.\\
\\
Lemma \ref{thm:main_spectral} bounds the spectral norm of the matrix $(\mathcal{R}_{\tilde{\Omega}} - \mathcal{I})(\matZ)$ using the sample set $\tilde{\Omega}$. 
\begin{lemma}\label{thm:main_spectral}
Let $\matZ \in \mathbb{R}^{m\times n}$ be a fixed matrix. Let $\tilde{\Omega}$ be a sample set in (\ref{eqn:omega_tilde}). Then,
for any universal constant $c>1$, we have 
$$
\TNorm{\left(\mathcal{R}_{\tilde{\Omega}} - \mathcal{I}\right)\matZ} \leq 2\sqrt{\frac{c}{c_0}} \muTNorm{\matZ}
+\frac{c}{c_0}\muNorm{\matZ}
$$
holding with probability at least 
$$1 - (m+n)^{1-c}.$$
\end{lemma}
%
%
%
%
%
%
Next two results control the $\mu(\infty,2)$ and $\mu(\infty)$ norms of the projection of a matrix after random sampling.\\
\begin{lemma}\label{thm:mu2norm} 
Let $\matZ \in \mathbb{R}^{m\times n}$ be a fixed matrix. Let $\tilde{\Omega}$ be a sample set in (\ref{eqn:omega_tilde}). Then,
for any universal constant $c>2$, we have 
$$\muTNorm{(\mathcal{P}_T\mathcal{R}_{\tilde{\Omega}} - \mathcal{P}_T)\matZ} \leq \frac{1}{2}\left(\muTNorm{\matZ}+ \muNorm{\matZ}\right)
$$
holding with probability at least 
$$1 - (m+n)^{2-c}.$$
\end{lemma}
\begin{lemma} \label{thm:munorm}
Let $\matZ \in \mathbb{R}^{m\times n}$ be a fixed matrix. Let $\tilde{\Omega}$ be a sample set in (\ref{eqn:omega_tilde}). Then,
for any universal constant $c>3$, we have 
$$\muNorm{(\mathcal{P}_T\mathcal{R}_{\tilde{\Omega}} - \mathcal{P}_T)\matZ} \leq \frac{1}{2}\muNorm{\matZ}
$$
holding with probability at least 
$$1 - (m+n)^{3-c}.$$
\end{lemma}
%
%
We now validate Condition 2 in Proposition \ref{prop:main} using the $\matY$ constructed above.\\

\textbf{Bounding $\FNorm{\matU\matV^T - \mathcal{P}_T(\matY)}$}\\
\\
We set
$\Delta_k = \matU\matV^T - \mathcal{P}_T(\matW_k)$,
for $k=1,...,k_0$.
Then, from definition of $\matW_k$ we have 
$$
\Delta_k = (\mathcal{P}_T - \mathcal{P}_T\mathcal{R}_{\Omega_k}\mathcal{P}_T)\Delta_{k-1}.
$$
%
%
%
%
%
%
We used $\mathcal{P}_T(\matU\matV^T) = \matU\matV^T$ and $\mathcal{P}_T\mathcal{P}_T(\matX) = \mathcal{P}_T(\matX)$. 
%
%
%
%
%
Using the independence of $\Delta_{k-1}$ and $\Omega_k$,
\begin{eqnarray*}
\FNorm{\Delta_k} = \FNorm{(\mathcal{P}_T - \mathcal{P}_T\mathcal{R}_{\Omega_k}\mathcal{P}_T)\Delta_{k-1}}
%
\leq \OpNorm{\mathcal{P}_T - \mathcal{P}_T\mathcal{R}_{\Omega_k}\mathcal{P}_T}\FNorm{\Delta_{k-1}}.
\end{eqnarray*}
We can bound this by recursively applying Lemma \ref{thm:operator} with $\Omega_k$, for all $k$. Thus,
\begin{eqnarray*}
\FNorm{\mathcal{P}_T(\matY) - \matU\matV^T} = \FNorm{\Delta_{k_0}}= 
\left(\frac{1}{2}\right)^{k_0}\FNorm{\matU\matV^T}
%
%
%
%
%
\leq \sqrt{\frac{\varrho}{(m+n)^{15}}} 
\end{eqnarray*}
The above result fails with probability at most $(m+n)^{1-c}$ for each $k$; thus, total probability of failure is at most $11(m+n)^{1-c}\text{ log}(m+n)$.\\

\textbf{Bounding $\TNorm{\mathcal{P}_{T^\perp}(\matY)}$}\\
\\
By definition, $\matY$ can be written as 
$$
\matY = \sum_{k=1}^{k_0} \mathcal{R}_{\Omega_k}\mathcal{P}_T\left(\matU\matV^T - \mathcal{P}_T(\matW_{k-1})\right)
= \sum_{k=1}^{k_0} \mathcal{R}_{\Omega_k}\mathcal{P}_T(\Delta_{k-1})
$$
It follows that,
\begin{eqnarray*}
\TNorm{\mathcal{P}_{T^\perp}(\matY)} 
%
%
= \TNorm{\mathcal{P}_{T^\perp}\sum_{k=1}^{k_0} \left(\mathcal{R}_{\Omega_k}\mathcal{P}_T - \mathcal{P}_T\right)(\Delta_{k-1})}
%
%
%
\leq \sum_{k=1}^{k_0} \TNorm{(\mathcal{R}_{\Omega_k}-I)(\Delta_{k-1})}
\end{eqnarray*}
We use
$$%
{\mathcal{P}_{T}(\Delta_{k})} = \mathcal{P}_{T}(\matU\matV^T - \mathcal{P}_T(\matW_k)) = \matU\matV^T - \mathcal{P}_T(\matW_k) = \Delta_k, \text{ for all } k.
$$
We apply Lemma \ref{thm:main_spectral} to each summand in the above inequality, with corresponding $\Omega_k$, to obtain 
\begin{eqnarray}\label{eqn:Pt_perp}
\TNorm{\mathcal{P}_{T^\perp}(\matY)} \leq 2\sqrt{\frac{c}{c_0}}\sum_{k=1}^{k_0}\muTNorm{\Delta_{k-1}}+ \frac{c}{{c_0}}\sum_{k=1}^{k_0}\muNorm{\Delta_{k-1}}
\end{eqnarray}
%
%
%
%
We can derive the following, applying Lemma \ref{thm:munorm} $k$ times, with $\Omega_k$, 
\begin{eqnarray}\label{eqn:Delta_mu}
\muNorm{\Delta_{k}} = \muNorm{(\mathcal{P}_T - \mathcal{P}_T\mathcal{R}_{\Omega_{k}})\Delta_{k-1}} \leq \left(\frac{1}{2}\right)^{i}\muNorm{\Delta_{k-i}} \leq \left(\frac{1}{2}\right)^{k}\muNorm{\matU\matV^T}
\end{eqnarray}
holding with failure probability at most $k\cdot(m+n)^{3-c}$, for all $k$.
\\
Similarly, applying Lemma \ref{thm:mu2norm} and Lemma \ref{thm:munorm} recursively, with $\Omega_k$, we can derive,
%
%
%
%
%
%
%
%
%
%
%
%
%
\begin{eqnarray}\label{eqn:Delta_mu2}
\nonumber \muTNorm{\Delta_k} 
&=& \muTNorm{(\mathcal{P}_T - \mathcal{P}_T\mathcal{R}_{\Omega_{k}}\mathcal{P}_T)\Delta_{k-1}}
\leq \frac{1}{2}\muNorm{\Delta_{k-1}} + \frac{1}{2}\muTNorm{\Delta_{k-1}}\\
\nonumber
\nonumber
\text{(step $j$)} &\leq& \sum_{i=1}^j\left(\frac{1}{2}\right)^i\muNorm{\Delta_{k-i}} +  \left(\frac{1}{2}\right)^j\muTNorm{\Delta_{k-j}}\\
\nonumber
&\leq& \sum_{i=1}^j\left(\frac{1}{2}\right)^i\left(\frac{1}{2}\right)^{k-i}\muNorm{\matU\matV^T} +  \left(\frac{1}{2}\right)^j\muTNorm{\Delta_{k-j}}\\
\nonumber
&\leq& j\left(\frac{1}{2}\right)^{k}\muNorm{\matU\matV^T} +  \left(\frac{1}{2}\right)^j\muTNorm{\Delta_{k-j}}\\
\text{(step $k$)} &\leq& k\left(\frac{1}{2}\right)^{k}\muNorm{\matU\matV^T} +  \left(\frac{1}{2}\right)^k\muTNorm{\matU\matV^T}
\end{eqnarray}
holding with failure probability at most $k\cdot(m+n)^{2-c}$, for all $k$.
Using (\ref{eqn:Delta_mu}) and (\ref{eqn:Delta_mu2}), it follows from (\ref{eqn:Pt_perp}),
\begin{eqnarray*}
\TNorm{\mathcal{P}_{T^\perp}(\matY)} 
%
&\leq& 2\sqrt{\frac{c}{c_0}}\sum_{k=1}^{k_0}(k-1)\left(\frac{1}{2}\right)^{k-1}\muNorm{\matU\matV^T} + 2\sqrt{\frac{c}{c_0}}\sum_{k=1}^{k_0} \left(\frac{1}{2}\right)^{k-1}\muTNorm{\matU\matV^T}\\
&& + \frac{c}{{c_0}}\sum_{k=1}^{k_0}\left(\frac{1}{2}\right)^{k-1}\muTNorm{\matU\matV^T}
\end{eqnarray*}
We note that, for all $(i,j)$,
$$
\abs{(\matU\matV^T)_{ij}} = \abs{\e_i^T\matU\matV^T\e_j}\leq \sqrt{\frac{\mu_i\varrho}{m}}\sqrt{\frac{\nu_j\varrho}{n}}\leq 1,
$$
$$
\TNorm{(\matU\matV^T)_{i,*}} = \TNorm{\e_i^T\matU\matV^T} = \sqrt{\frac{\mu_i\varrho}{m}}, \qquad 
\TNorm{(\matU\matV^T)_{*,j}} = \TNorm{\matU\matV^T\e_j} = \sqrt{\frac{\nu_j\varrho}{n}} 
$$
Thus,
\begin{eqnarray*}
\muTNorm{\matU\matV^T}
&=& \max\left\{\max_i\sqrt{\frac{m}{\mu_i \varrho}}\TNorm{(\matU\matV^T)_{i,*}},\max_j\sqrt{\frac{n}{\nu_j \varrho}}\TNorm{(\matU\matV^T)_{*,j}}\right\}
=1
\end{eqnarray*}
Therefore,
\begin{eqnarray}
\nonumber 
\TNorm{\mathcal{P}_{T^\perp}(\matY)} 
%
&\leq& 2\sqrt{\frac{c}{c_0}}\sum_{k=1}^{k_0}\left((k-1)\left(\frac{1}{2}\right)^{k-1}+ \left(\frac{1}{2}\right)^{k-1}\right)
 +\frac{c}{c_0}\sum_{k=1}^{k_0}\left(\frac{1}{2}\right)^{k-1}\\
%
%
\nonumber
&<& 2\sqrt{\frac{c}{c_0}}\sum_{k=1}^{\infty}k\left(\frac{1}{2}\right)^{k-1}
 +\frac{c}{c_0}\sum_{k=1}^{\infty}\left(\frac{1}{2}\right)^{k-1}
%
= 8\sqrt{\frac{c}{c_0}}
 +\frac{2c}{c_0} \leq \frac{1}{2},
\end{eqnarray}
%
%
%
by setting $c_0 \geq  {264c}$.

Let, the total number of sampled entries be $s$. Expected number of observed entries required to solve (\ref{eqn:main_problem}) is 
$\mathbb{E}(s) = \sum_{i,j}p_{ij} 
=  O{(((m+n)\varrho - \varrho^2)\log^2 (m+n))}.
$
Summing up the individual failure probabilities of Proposition \ref{prop:main} the total failure probability never exceeds $33\cdot \log(m+n)(m+n)^{3-c}$, for sufficiently large $c>3$.

Finally, we can apply Hoeffding's inequality to show that $s$ is sharply concentrated around its expectation, i.e., 
$s = \Theta(((m+n)\varrho - \varrho^2)\log^2 (m+n))$ 
with probability at least $1 - 66\text{ log}(m+n)(m+n)^{3-c}$, for sufficiently large $c>3$.\\
%
%

This completes the proof of Theorem \ref{thm:main}.

\section{Proof of Theorem \ref{thm:row_coherent}}\label{sec:thm2}
We closely follow the proof given by \cite{BCSW14}. We pick each row of $\matM$ with some probability $p$ and observe all the entries of this sampled row. 
Let $\Gamma \subseteq [m]$ be the set of indices of the row picked, and 
$\mathcal{S}_\Gamma(\matX)$ be a matrix obtained from $\matX$ by zeroing out the rows outside $\Gamma$. 
Recall, SVD of $\matM$ is $\matU\matSig\matV^T$. We use the following lemma (Lemma 14 of \cite{BCSW14}).
\begin{lemma}\label{lem:row_sample}
Let $\mu_i \leq \max_i\frac{m}{\rho}\TsNorm{\matU^T\e_i}^2 \leq \mu_0$, $\forall i \in [m]$, and $p \geq c_2\frac{\mu_0\varrho}{m}\log m$ for some universal constant $c_2$. Then, for any universal constant $c>1$, and $c_2 \geq 20c$,
$$
\TNorm{\matU^T\mathcal{S}_\Gamma(\matU) - \matI_{\varrho}} \leq {1}/{2},
$$
holds with probability at least $1-(m+n)^{1-c}$, 
where $\matI_\varrho$ is the identity matrix in $\mathbb{R}^{\varrho \times \varrho}$.
\end{lemma}

Now, $\TNorm{\matU^T\mathcal{S}_\Gamma(\matU) - \matI_{\varrho}} \leq {1}/{2}$ implies that $\matU^T\mathcal{S}_\Gamma(\matU)$ is invertible and $\mathcal{S}_\Gamma(\matU) \in \mathbb{R}^{m \times \varrho}$ has rank-$\varrho$. Using SVD of $\matM$, we can write 
$\mathcal{S}_\Gamma(\matM) = \mathcal{S}_\Gamma(\matU)\matSig\matV^T$, and this has full rank-$\varrho$. Therefore, $\mathcal{S}_\Gamma(\matM)$ and $\matM$ have the same row space, and we conclude that $\tilde{\nu}_j = \nu_j$, $\forall j \in [n]$. Thus, using the sample set $\Omega$ in Algorithm \ref{alg:row_coherent} we can recover $\matM$ exactly via nuclear norm minimization in (\ref{eqn:main_problem}), with high probability. Expected number of entries observed in Algorithm \ref{alg:row_coherent} is 
\begin{eqnarray*}
pmn + \sum_{i,j}^{m,n}p_{ij} 
&=&
O(\mu_0((m+2n)\varrho - \varrho^2)\log^2(m+n)),
\end{eqnarray*}
where, $p_{ij}$ as in (\ref{eqn:pij_alg1}).
We apply standard Hoeffding inequality to bound the actual sample size, and Theorem \ref{thm:row_coherent} follows as a corollary of Theorem \ref{thm:main}.

%% file: Proofs.tex
\input{main_proof}
\section{Proof of Technical Lemmas}\label{sec:proof_lemmas}
Here we prove Lemmas \ref{thm:operator} through \ref{lem:row_sample} using the matrix Bernstein inequality of Lemma \ref{thm:matrix_bernstein} as the main tool. Also, we frequently use the fact in (\ref{eqn:proj_relax}) and the result in Lemma 
\ref{lemma:relax}.
Note that $\mathcal{P}_{T}$ is self-adjoint linear operator. Thus we can write the following for any $\matX \in \mathbb{R}^{m \times n}$:
\begin{eqnarray}\label{eqn:adjoint}
\mathcal{P}_T(\matX) = \sum_{i,j}\left<\mathcal{P}_T(\matX),\e_i\e_j^T\right>\e_i\e_j^T = \sum_{i,j}\left<\mathcal{P}_T(\matX),\mathcal{P}_T(\e_i\e_j^T)\right>\e_i\e_j^T= \sum_{i,j}\left<\matX,\mathcal{P}_T(\e_i\e_j^T)\right>\e_i\e_j^T
\end{eqnarray}
We can derive, for all $i$ and $j$,
\begin{eqnarray}\label{eqn:proj_relax}
\FNormS{\mathcal{P}_T\left(\e_i\e_j^T\right)} =  \left<\mathcal{P}_T\left(\e_i\e_j^T\right), \e_i\e_j^T\right> 
=\frac{\mu_i \varrho}{m} + \frac{\nu_j \varrho}{n} - \frac{\mu_i \varrho}{m}\cdot \frac{\nu_j \varrho}{n}
\end{eqnarray}
%
%
%
%
Also, we know for all $i$, $j$,
\begin{eqnarray}\label{eqn:lev_relation}
0\leq \frac{\mu_i\varrho}{m} \leq \sqrt{\frac{\mu_i\varrho}{m}} \leq 1, 
\quad 0\leq \frac{\nu_j\varrho}{n} \leq \sqrt{\frac{\nu_j\varrho}{n}}\leq 1.
\end{eqnarray}
\begin{lemma}\label{lemma:relax}
Using our notations, for all $i$, $j$,
$$
\frac{\mu_i\varrho}{m}+\frac{\nu_j\varrho}{n} - \frac{\mu_i\varrho}{m}\cdot \frac{\nu_j\varrho}{n}
\geq \sqrt{\frac{\mu_i\varrho}{m}}\cdot \sqrt{\frac{\nu_j\varrho}{n}}
\geq \frac{\mu_i\varrho}{m}\cdot \frac{\nu_j\varrho}{n}
$$
\end{lemma}
\begin{Proof}
Let, $x = \frac{\mu_i\varrho}{m}$ and $y=\frac{\nu_j\varrho}{n}$. Then,
\begin{eqnarray*}
(x+y-xy)^2 
&=&  xy + (x^2-x^2y) + (y^2-xy^2) + x^2y^2 + xy - x^2y - xy^2\\
&=& xy + x^2(1-y) + y^2(1-x) + xy(1 - x)(1 - y)\\
&\geq& xy \quad \text{using } (\ref{eqn:lev_relation})
\end{eqnarray*}
Also, $x+y-xy \geq 0$. Thus, $x+y-xy \geq \sqrt{xy} \geq xy$. 
The last inequality follows because $0 \leq x,y \leq 1$.
\end{Proof}

\input{main_tool}
We consider sampling probabilities $\{q_{ij}\}$ of the form (\ref{eqn:probability_qij}) to prove Lemmas \ref{thm:operator} through \ref{thm:munorm}. \\

\textbf{Notation Overloading:} For simplicity, we reuse some of the notations in Section \ref{sec:thm_operator} through \ref{sec:thm_munorm}. Specifically, we replace $\tilde{\Omega}$ by $\Omega$ to denote a sample set in (\ref{eqn:omega_tilde}),
and, $\delta_{ij} = \mathbb{I}((i,j)\in \tilde{\Omega})$.
\subsection{Proof of Lemma \ref{thm:operator}}\label{sec:thm_operator}
\input{lemma_9_our_modified}
\subsection{Proof of Lemma \ref{thm:main_spectral}}
\input{lemma_10_our_modified}
\subsection{Proof of Lemma \ref{thm:mu2norm}}
\input{lemma_11_our_modified}
\subsection{Proof of Lemma \ref{thm:munorm}}\label{sec:thm_munorm}
\input{lemma_12_our_modified}
\input{theorem_2_our_modified}
%

%% file: main_proof.tex
\section{Proof of Optimality Conditions in Proposition \ref{prop:main}}\label{sec:optimality}
Let $\mat\matM$ be the low-rank target matrix with rank-$\varrho$ SVD $\matM = \matU\matSig \matV^T$. 
We want to show that any perturbation $\matZ$ to $\matM$, such that, $\matM + \matZ$ is a solution to (\ref{eqn:main_problem}), strictly increases the nuclear norm, unless $\matZ=0$. Now, $\matM + \matZ$ is feasible only if $\mathcal{P}_\Omega(\matM + \matZ) = \mathcal{P}_\Omega(\matM)$, which implies  
$\mathcal{R}_{\Omega}(\matZ) = 0$, e.g., $\matZ$ is in the null space of $\mathcal{R}_{\Omega}$ operator. 
We can choose $\matU_{\perp}$ and $\matV_{\perp}$ such that $[\matU, \matU_{\perp}]$ and $[\matV, \matV_{\perp}]$ are unitary matrices for which $\left<\matU_{\perp}\matV_{\perp}^T,\mathcal{P}_{T^\perp}(\matZ)\right> = \NNorm{\mathcal{P}_{T^\perp}(\matZ)}$. 
Then it follows from standard inequality of trace norm, for some $\matY$ in the range of $\mathcal{R}_{\Omega}$,
\begin{eqnarray*}
\NNorm{\matM+\matZ} 
&\geq& \left<\matU\matV^T + \matU_{\perp}\matV_{\perp}^T,\matM+\matZ\right> \\
&=& \NNorm{\matM} + \left<\matU\matV^T + \matU_{\perp}\matV_{\perp}^T,\matZ\right> \\
%
%
& = & \NNorm{\matM}+\left<\matU\matV^T - \mathcal{P}_T(\matY),\mathcal{P}_{T}(\matZ)\right> + \left<\matU_\perp\matV_\perp^T - \mathcal{P}_{T^\perp}(\matY),\mathcal{P}_{T^\perp}(\matZ)\right>\\
& \overset{(a)}{\geq} & \NNorm{\matM} - \FsNorm{\matU\matV^T - \mathcal{P}_T(\matY)}\cdot \FsNorm{\mathcal{P}_{T}(\matZ)}+ \NNorm{\mathcal{P}_{T^\perp}(\matZ)} - \left<\mathcal{P}_{T^\perp}(\matY),\mathcal{P}_{T^\perp}(\matZ)\right>\\
& \geq & \NNorm{\matM} -  \FsNorm{\matU\matV^T - \mathcal{P}_T(\matY)}\cdot \FsNorm{\mathcal{P}_{T}(\matZ)}
+ \left(1 - \TNorm{\mathcal{P}_{T^\perp}(\matY)}\right)\NNorm{\mathcal{P}_{T^\perp}(\matZ)}\\
\nonumber 
& \overset{(b)}{\geq} & \NNorm{\matM} + \left(1 - \TNorm{\mathcal{P}_{T^\perp}(\matY)} -  \frac{\FsNorm{\matU\matV^T - \mathcal{P}_T(\matY)}\left(\max_{i,j}\frac{1}{\sqrt{p_{ij}}}\right)}{{\left(1- \OpNorm{\mathcal{P}_T\mathcal{R}_{\Omega}\mathcal{P}_T - \mathcal{P}_T}\right)}^{\frac{1}{2}}}\right)\NNorm{\mathcal{P}_{T^\perp}(\matZ)}\\%
&>& \NNorm{\matM}
\end{eqnarray*}
Above, $(a)$ follows from Von-Neumann trace inequality, and (b) follows from Lemma \ref{thm:sqrt_pij}. Using $\max_{i,j}\frac{1}{\sqrt{p_{ij}}} \leq (mn)^{5/2}\leq (m+n)^5$, and the conditions in Proposition \ref{prop:main}, we derive the final inequality. Note that, Condition 1 in Proposition \ref{prop:main} implies $\mathcal{R}_\Omega$ is the identity operator on the elements of subspace $T$, therefore $\mathcal{P}_{T^\perp}(\matZ) =0$ implies $\matZ = 0$.
%
%
\\%

The following lemma is similar to Lemma 13 of \cite{BCSW14}.
\begin{lemma}\label{thm:sqrt_pij}
For any $\matZ \in \mathbb{R}^{m \times n}$, s.t., $\mathcal{P}_{\Omega}(\matZ)=0$,
$$
\FNorm{\mathcal{P}_T(\matZ)}\leq {\left(1- \OpNorm{\mathcal{P}_T\mathcal{R}_{\Omega}\mathcal{P}_T - \mathcal{P}_T}\right)}^{-\frac{1}{2}}\left(\max_{i,j}\frac{1}{\sqrt{p_{ij}}}\right)\NNorm{\mathcal{P}_{T^\perp}\matZ}
$$
%
\end{lemma}
\begin{Proof}
Let us define the operator $\mathcal{R}_{\Omega}^{1/2}: \mathbb{R}^{m \times n} \rightarrow \mathbb{R}^{m \times n}$ as
$$\mathcal{R}_{\Omega}^{1/2}(\matZ) := \sum_{i,j}\frac{1}{\sqrt{p_{ij}}}\delta_{ij}\left<\e_i\e_j^T,\matZ\right>\e_i\e_j^T$$
Note that $\mathcal{R}_{\Omega}^{1/2}$ is self-adjoint, and $\mathcal{R}_{\Omega}^{1/2}\mathcal{R}_{\Omega}^{1/2}=\mathcal{R}_{\Omega}$. Therefore, we have
\begin{eqnarray}\label{eqn:R_oemga_pt_Z}
\nonumber
\FNormS{\mathcal{R}_{\Omega}^{1/2}\mathcal{P}_T(\matZ)} 
&=& {\left<\mathcal{R}_{\Omega}\mathcal{P}_T(\matZ), \mathcal{P}_T(\matZ)\right>}\\
\nonumber
&=& {\left<\mathcal{P}_T\mathcal{R}_{\Omega}\mathcal{P}_T(\matZ), \mathcal{P}_T(\matZ)\right>}\\
\nonumber
%
&=& {\left<\mathcal{P}_T\mathcal{R}_{\Omega}\mathcal{P}_T(\matZ) - \mathcal{P}_T(\matZ), \mathcal{P}_T(\matZ)\right> + \left<\mathcal{P}_T(\matZ), \mathcal{P}_T(\matZ)\right>}\\
%
%
%
&\geq& {(1- \OpNorm{\mathcal{P}_T\mathcal{R}_{\Omega}\mathcal{P}_T - \mathcal{P}_T})}\cdot \FNormS{ \mathcal{P}_T(\matZ)}
\end{eqnarray}
%
%
%
%
%
%
%
%
%
Also, we have $\FNorm{\mathcal{R}_{\Omega}^{1/2}(\matZ)} = 0$ for any $\matZ$ s.t. $\mathcal{P}_{\Omega}(\matZ)=0$.
It follows,
\begin{eqnarray}\label{eqn:R_oemga_pt_Z2}
\nonumber
0 = \FNorm{\mathcal{R}_{\Omega}^{1/2}(\matZ)} &\geq& \FNorm{\mathcal{R}_{\Omega}^{1/2}\mathcal{P}_T(\matZ)} - \FNorm{\mathcal{R}_{\Omega}^{1/2}\mathcal{P}_{T^\perp}(\matZ)}\\
%
 \quad \FNorm{\mathcal{R}_{\Omega}^{1/2}\mathcal{P}_T(\matZ)} &\leq& \FNorm{\mathcal{R}_{\Omega}^{1/2}\mathcal{P}_{T^\perp}(\matZ)}
\leq \left(\max_{i,j}\frac{1}{\sqrt{p_{ij}}}\right)\FNorm{\mathcal{P}_{T^\perp}(\matZ)},
\end{eqnarray}
where we use
$$\FNorm{\mathcal{R}_{\Omega}^{1/2}\mathcal{P}_{T^\perp}(\matZ)} \leq \max_{i,j}\frac{1}{\sqrt{p_{ij}}}\FNorm{\sum_{i,j}\delta_{ij}\left<\e_i\e_j^T,\mathcal{P}_{T^\perp}(\matZ)\right>\e_i\e_j^T} \leq \max_{i,j}\frac{1}{\sqrt{p_{ij}}}\FNorm{\mathcal{P}_{T^\perp}(\matZ)}$$
%
%
Combining (\ref{eqn:R_oemga_pt_Z}) and (\ref{eqn:R_oemga_pt_Z2}), and using $\FNorm{\matX}\leq \NNorm{\matX}$,
\begin{eqnarray*}
\sqrt{(1- \OpNorm{\mathcal{P}_T\mathcal{R}_{\Omega}\mathcal{P}_T - \mathcal{P}_T})}\cdot \FNorm{ \mathcal{P}_T(\matZ)}
\leq \left(\max_{i,j}\frac{1}{\sqrt{p_{ij}}}\right)\FNorm{\mathcal{P}_{T^\perp}(\matZ)}
\leq \left(\max_{i,j}\frac{1}{\sqrt{p_{ij}}}\right)\NNorm{\mathcal{P}_{T^\perp}(\matZ)}
\end{eqnarray*}
The result follows.
%
%
%
%
\end{Proof}
%
%

%% file: main_tool.tex
%
\begin{lemma}\label{thm:matrix_bernstein}
(\cite{Tropp12}, [Theorem 16] of \cite{BCSW14})\\
Let $\matX_1$, ..., $\matX_N \in \mathbb{R}^{m \times n}$ be independent, zero-mean random matrices. Suppose
$$
\max\left\{ \TNorm{\sum_{t=1}^{N}\mathbb{E}\left[\matX_t\matX_t^T\right]}, \TNorm{\sum_{t=1}^{N}\mathbb{E}\left[\matX_t^T\matX_t\right]}\right\} \leq \sigma^2
$$
and 
$
\TNorm{\matX_t} \leq \gamma
$
almost surely for all $t$. Then for any $c>0$, we have
\begin{eqnarray*}
\TNorm{\sum_{t=1}^{N}\matX_t} \leq 2\sqrt{c \sigma^2\log(m+n)} + c \gamma \log(m+n)
\end{eqnarray*}
with probability at least $1 - (m+n)^{-(c-1)}$.
\end{lemma}

%% file: lemma_9_our_modified.tex
%
%
%
For any matrix $\matZ \in \mathbb{R}^{m \times n}$, we can write  
%
%
%
%
%
%
%
%
%
\begin{eqnarray*}
 \left(\mathcal{P}_T\mathcal{R}_{\Omega}\mathcal{P}_T - \mathcal{P}_T\right)(\matZ) 
%
=  \sum_{i,j}\left(\frac{1}{q_{ij}}\delta_{ij}-1\right)\left<\mathcal{P}_T\left(\e_i\e_j^T\right),\matZ\right>\mathcal{P}_T\left(\e_i\e_j^T\right) 
= \sum_{i,j}\mathcal{S}_{ij}(\matZ).
\end{eqnarray*}
Using $\mathbb{E}\left[\delta_{ij}\right] = q_{ij}$, we have $\mathbb{E}[\mathcal{S}_{ij}(\matZ)]=0$ for any $\matZ$. Thus, we conclude that $\mathbb{E}[\mathcal{S}_{ij}] = 0$. Also, $\mathcal{S}_{ij}$'s are independent of each other. Using probabilities in (\ref{eqn:probability_qij}) ($\mathcal{S}_{ij}$'s vanish when $q_{ij}=1$, for all $\matZ$ and $(i,j)$), and (\ref{eqn:proj_relax}), we derive
%
%
%
%
\begin{eqnarray*}
\FNorm{\mathcal{S}_{ij}(\matZ)} 
\leq \frac{1}{q_{ij}}\FNormS{\mathcal{P}_T\left(\e_i\e_j^T\right)}\FNorm{\matZ}
\leq \frac{\FNorm{\matZ}}{c_0\cdot \log(m+n)}.
\end{eqnarray*}
From definition of operator norm, $\OpNorm{\mathcal{S}_{ij}} \leq \frac{1}{c_0\cdot \text{log}(m+n)}$. Also, we derive 
\begin{eqnarray*}
\mathbb{E}\left[\mathcal{S}^2_{ij}(\matZ)\right] &=&\mathbb{E}\left[\left(\frac{1}{q_{ij}}\delta_{ij}-1\right)^2\right]\left<\e_i\e_j^T,\mathcal{P}_T(\matZ)\right>\left<\e_i\e_j^T,\mathcal{P}_T(\e_i\e_j^T)\right>\mathcal{P}_T\left(\e_i\e_j^T\right)\\
&=&\frac{1-q_{ij}}{q_{ij}}\left<\e_i\e_j^T,\mathcal{P}_T(\matZ)\right>\left<\e_i\e_j^T,\mathcal{P}_T(\e_i\e_j^T)\right>\mathcal{P}_T\left(\e_i\e_j^T\right)
%
%
\end{eqnarray*}
%
\begin{eqnarray*}
\nonumber
\FNorm{\sum_{i,j}\mathbb{E}\left[\mathcal{S}^2_{ij}(\matZ)\right]} 
&\leq&\left(\max_{i,j}\frac{1-q_{ij}}{q_{ij}}\FNormS{\mathcal{P}_T(\e_i\e_j^T)}\right)\FNorm{\sum_{i,j}\left<\e_i\e_j^T,\mathcal{P}_T(\matZ)\right>\mathcal{P}_T\left(\e_i\e_j^T\right)}\\
\nonumber &=&\left(\max_{i,j}\frac{1-q_{ij}}{q_{ij}}\FNormS{\mathcal{P}_T(\e_i\e_j^T)}\right)\FNorm{\mathcal{P}_T\left(\sum_{i,j}\left<\e_i\e_j^T,\mathcal{P}_T(\matZ)\right>\left(\e_i\e_j^T\right)\right)}\\
%
%
\nonumber &=&\left(\max_{i,j}\frac{1-q_{ij}}{q_{ij}}\FNormS{\mathcal{P}_T(\e_i\e_j^T)}\right)\FNorm{\mathcal{P}_T(\matZ)}
%
%
\end{eqnarray*}
%
\begin{eqnarray*}
\OpNorm{\sum_{i,j}\mathbb{E}\left[\mathcal{S}^2_{ij}\right]} &\leq& \max_{i,j}\frac{1-q_{ij}}{q_{ij}}\FNormS{\mathcal{P}_T(\e_i\e_j^T)} \quad \leq \frac{1}{c_0\cdot \text{log}(m+n)}
\end{eqnarray*}
We apply Matrix Bernstein inequality in Lemma \ref{thm:matrix_bernstein} 
using
$$\sigma^2 = \frac{1}{c_0\cdot \text{log}(m+n)}, \quad \gamma = \frac{1}{c_0\cdot \text{log}(m+n)},$$
to obtain, for any $c>1$, $c_0 \geq 20c$,
\begin{eqnarray*}
\OpNorm{\mathcal{P}_T\mathcal{R}_{\Omega}\mathcal{P}_T - \mathcal{P}_T} \leq {1}/{2}
\end{eqnarray*}
holding with probability at least
$$1 - (m+n)^{(1-c)}.$$
%
%

%% file: lemma_10_our_modified.tex
%
%
%
We can write the matrix $\left(\mathcal{R}_{\Omega} - \mathcal{I}\right)\matZ$ as sum of independent matrices:
$$\left(\mathcal{R}_{\Omega} - \mathcal{I}\right)\matZ = \sum_{i,j}\left(\frac{1}{q_{ij}}\delta_{ij}-1\right)\matZ_{ij}\e_i\e_j^T = \sum_{i,j}\matS_{ij}.$$
We note that, $\mathbb{E}[\matS_{ij}] = 0$, and $\matS_{ij}$'s are zero matrix when $q_{ij}=1$, for all $(i,j)$. We have
$\TNorm{\matS_{ij}} 
\leq \frac{\abs{\matZ_{ij}}}{q_{ij}}$.
Moreover,
\begin{eqnarray*}
\sum_{i,j}\mathbb{E}\left[\matS_{ij}\matS_{ij}^T\right] 
= \sum_{i,j}\matZ_{ij}^2\e_i\e_i^T\mathbb{E}\left[\left(\frac{1}{q_{ij}}\delta_{ij}-1\right)^2\right] 
= \sum_{i}\left(\sum_{j}\matZ_{ij}^2\frac{1-q_{ij}}{q_{ij}}\right)\e_i\e_i^T 
%
%
\end{eqnarray*}
Thus,
$$
\TNorm{\sum_{i,j}\mathbb{E}\left[\matS_{ij}\matS_{ij}^T\right]} 
\leq \max_i \sum_{j=1}^{n}\frac{1-q_{ij}}{q_{ij}}\matZ_{ij}^2 
$$
Similarly,
$$
\TNorm{\sum_{i,j}\mathbb{E}\left[\matS_{ij}^T\matS_{ij}\right]} 
\leq \max_j \sum_{i=1}^{m}\frac{1-q_{ij}}{q_{ij}}\matZ_{ij}^2 
$$
%
%
%
%
%
%
%
Clearly, when $q_{ij} = 1$ the above quantities are zero. 
Using $q_{ij}$ in (\ref{eqn:probability_qij}), and Lemma \ref{lemma:relax}, we have
%
\begin{eqnarray*}
\TNorm{\matS_{ij}}
\leq \frac{1}{c_0\cdot \text{log}(m+n) }\abs{\matZ_{ij}}\sqrt{\frac{m}{\mu_i \varrho}}\sqrt{\frac{n}{\nu_j \varrho}} 
\leq \frac{\muNorm{\matZ}}{c_0\cdot \text{log}(m+n) }.
\end{eqnarray*}
Using $q_{ij}$ in (\ref{eqn:probability_qij}), and noting that 
$
\left(\frac{\mu_i\varrho}{m}+\frac{\nu_j\varrho}{n}-\frac{\mu_i\varrho}{m}\cdot \frac{\nu_j\varrho}{n}\right) \geq \frac{\mu_i\varrho}{m}
$, we have
\begin{eqnarray*}
\sum_{j=1}^{n}\frac{1-q_{ij}}{q_{ij}}\matZ_{ij}^2&\leq& \frac{1}{c_0\cdot \text{log}(m+n)}\cdot \frac{m}{\mu_i \varrho}\sum_{j=1}^{n}\matZ_{ij}^2
\leq \frac{\muTNorm{\matZ}^2}{c_0\cdot \text{log}(m+n)}.
\end{eqnarray*}
Similarly, 
%
$$
\sum_{i=1}^{m}\frac{1-q_{ij}}{q_{ij}}\matZ_{ij}^2 
\leq \frac{1}{c_0\cdot \text{log}(m+n)}\cdot \frac{n}{\nu_j \varrho}\sum_{i=1}^{m}\matZ_{ij}^2 
\leq \frac{\muTNorm{\matZ}^2}{c_0\cdot \text{log}(m+n)}.
$$
%
The lemma follows from Matrix Bernstein inequality in Lemma \ref{thm:matrix_bernstein}, with  
$$
\gamma\text{ log}(m+n) \leq \frac{1}{c_0}\muNorm{\matZ}, 
\quad
\sigma^2 \text{ log}(m+n)
\leq \frac{1}{c_0}\muTNorm{\matZ}^2.
$$
%
%
%

%% file: lemma_11_our_modified.tex
%
Let, $$\matX = (\mathcal{P}_T\mathcal{R}_{\Omega} - \mathcal{P}_T)\matZ = \sum_{i,j}\left(\frac{\delta_{ij}}{q_{ij}}-1\right)\matZ_{ij}\mathcal{P}_T(\e_i\e_j^T)$$
%
%
Weighted $b$-th column of $\matX$ can be written as sum of independent, zero-mean column vectors.
%
$$
\sqrt{\frac{n}{\nu_b\varrho}}\matX_{*, b} 
= \sum_{i,j}\left(\frac{\delta_{ij}}{q_{ij}}-1\right)\matZ_{ij}\left(\mathcal{P}_T(\e_i\e_j^T)\e_b\right)\sqrt{\frac{n}{\nu_b\varrho}} 
= \sum_{i,j}\s_{ij}
$$
Clearly, $\mathbb{E}[\s_{ij}] = 0$.
We need bounds on $\TNorm{\s_{ij}}$ and $\TNorm{\sum_{i,j}\mathbb{E}\left[\s_{ij}^T\s_{ij}\right]}$ to apply Matrix Bernstein inequality.
First, we need to bound $\TsNorm{\mathcal{P}_T(\e_i\e_j^T)\e_b}$.
%
%
%
%
\begin{eqnarray}\label{lemma3_cases}
\nonumber
\TsNorm{\mathcal{P}_T(\e_i\e_j^T)\e_b} 
%
&=& \TsNorm{\matU\matU^T(\e_i\e_j^T)\e_b + (\e_i\e_j^T)\matV\matV^T\e_b - \matU\matU^T(\e_i\e_j^T)\matV\matV^T\e_b}\\
&=&
\begin{cases}
\TsNorm{\matU\matU^T\e_i + \left(\matI - \matU\matU^T\right)\e_i\TsNorm{\matV^T\e_b}^2} \leq \sqrt{\frac{\mu_i\varrho}{m}} + {\frac{\nu_b\varrho}{n}} & j=b,\\
\TsNorm{\left(\matI - \matU\matU^T\right)\e_i {\e_j^T\matV\matV^T\e_b}} \leq \abs{\e_j^T\matV\matV^T\e_b} & j \neq b,
\end{cases}
\end{eqnarray}
%
%
%
%
%
%
%
Above we use triangle inequality and definition of $\mu_i$ and $\nu_b$.
Note that, $\s_{ij}$ is a zero vector when $q_{ij}=1$, for all $(i,j)$. 
%
Otherwise, for $q_{ij}\neq 1$, we consider two cases. Using bounds in (\ref{lemma3_cases}), we have for $j=b$,
\begin{eqnarray*}
\TNorm{\s_{ij}} \leq \frac{1}{q_{ib}}\abs{\matZ_{ib}}\sqrt{\frac{n}{\nu_b\varrho}}\left(\sqrt{\frac{\mu_i\varrho}{m}} + \frac{\nu_b\varrho}{n}\right)
\end{eqnarray*}
Using $q_{ij}$ in (\ref{eqn:probability_qij}), 
$ 
q_{ib} \geq c_0\text{ log}(m+n)\sqrt{\frac{\mu_i\varrho}{m}}\sqrt{\frac{\nu_b\varrho}{n}}
$
and
$
q_{ib} \geq c_0\text{ log}(m+n)\cdot {\frac{\mu_i\varrho}{m}}
$.
%
%
%
Combining these two inequalities, we have 
%
\begin{eqnarray*}
\TNorm{\s_{ij}} \text{ log}(m+n)
%
\leq \frac{2}{c_0}\abs{\matZ_{ib}}\sqrt{\frac{m}{\mu_i\varrho}}\cdot \sqrt{\frac{n}{\nu_b\varrho}}\frac{\left(\sqrt{\frac{\mu_i\varrho}{m}} + \frac{\nu_b\varrho}{n}\right)}{\left(\sqrt{\frac{\mu_i\varrho}{m}} + \sqrt{\frac{\nu_b\varrho}{n}}\right)}
%
%
\leq \frac{2}{c_0}\muNorm{\matZ}
%
\end{eqnarray*}
%
%
%
For $j\neq b$, using 
$ 
q_{ib} \geq c_0\text{ log}(m+n)\sqrt{\frac{\mu_i\varrho}{m}}\sqrt{\frac{\nu_b\varrho}{n}}
$
(Lemma \ref{lemma:relax})
and $\abs{\e_j^T\matV\matV^T\e_b}  \leq \sqrt{\frac{\nu_j\varrho}{n}\cdot \frac{\nu_b\varrho}{n}}$,
%
%
\begin{eqnarray*}
\TNorm{\s_{ij}} 
%
%
\leq \frac{1}{q_{ij}}\abs{\matZ_{ij}}\sqrt{\frac{n}{\nu_b\varrho}} \cdot \sqrt{\frac{\nu_j\varrho}{n}}\cdot \sqrt{\frac{\nu_b\varrho}{n}}
\leq \frac{2}{c_0\text{ log}(m+n)}\muNorm{\matZ}
%
\end{eqnarray*}
Therefore, for all $(i,j)$, we have $\TNorm{\s_{ij}} \leq \frac{2}{c_0\text{ log}(m+n)}\muNorm{\matZ}$ .\\
%
%
On the other hand,
\begin{eqnarray*}
\abs{\sum_{i,j}\mathbb{E}\left[\s_{ij}^T\s_{ij}\right]} 
%
%
&=& \left(\sum_{j=b,i}+\sum_{j\neq b,i}\right)\frac{1-q_{ij}}{q_{ij}}\matZ_{ij}^2\TNormS{\mathcal{P}_T(\e_i\e_j^T)\e_b}\cdot {\frac{n}{\nu_b\varrho}}
\end{eqnarray*}
The above quantity is zero for $q_{ij}=1$. Otherwise, for $q_{ij}\neq 1$, we consider two cases.
\\
For $j=b$, using (\ref{lemma3_cases}) we have,
$
\TNormS{\mathcal{P}_T(\e_i\e_j^T)\e_b} \leq \left(\sqrt{\frac{\mu_i\varrho}{m}} + \sqrt{\frac{\nu_b\varrho}{n}}\right)^2
\leq 2\left({\frac{\mu_i\varrho}{m}} + {\frac{\nu_b\varrho}{n}}\right)
$.\\
%
Using $q_{ij}$ in (\ref{eqn:probability_qij}),
%
we have,
\begin{eqnarray*}
\sum_{j=b,i}
%
\leq 2\sum_{i}\frac{1-q_{ib}}{q_{ib}}\matZ_{ib}^2\left({\frac{\mu_i\varrho}{m}} + {\frac{\nu_b\varrho}{n}}\right)\cdot {\frac{n}{\nu_b\varrho}}
%
%
\leq \frac{4}{c_0\text{ log}(m+n)} \muTNorm{\matZ}^2,
\end{eqnarray*}
where we use the following bound in the second inequality. For all $(i,j)$, $q_{ij}\neq 0$,
\begin{eqnarray*}
\frac{\frac{\mu_i\varrho}{m}+\frac{\nu_j\varrho}{n}}{\frac{\mu_i\varrho}{m}+\frac{\nu_j\varrho}{n} - \frac{\mu_i\varrho}{m}\cdot\frac{\nu_j\varrho}{n}} 
=
1+\frac{\frac{\mu_i\varrho}{m}\cdot\frac{\nu_j\varrho}{n}}{\frac{\mu_i\varrho}{m}+\frac{\nu_j\varrho}{n} - \frac{\mu_i\varrho}{m}\cdot\frac{\nu_j\varrho}{n}}
\leq 1+\frac{\frac{\mu_i\varrho}{m}\cdot\frac{\nu_j\varrho}{n}}{\max\{\frac{\mu_i\varrho}{m},\frac{\mu_j\varrho}{n}\}}
%
%
\leq 2.
\end{eqnarray*}
For $j\neq b$, using 
$q_{ij} \geq c_0\text{ log}(m+n) \cdot \frac{\mu_j\varrho}{n}$ 
and (\ref{lemma3_cases}),
\begin{eqnarray*}
\sum_{j\neq b,i}
%
&\leq& \sum_{j\neq b,i}\frac{1-q_{ij}}{q_{ij}}\matZ_{ij}^2\abs{\e_j^T\matV\matV^T\e_b}^2\cdot {\frac{n}{\nu_b\varrho}}\\
&=& {\frac{n}{\nu_b\varrho}}\sum_{j\neq b}\abs{\e_j^T\matV\matV^T\e_b}^2\sum_{i}\frac{1-q_{ij}}{q_{ij}}\matZ_{ij}^2\\
%
%
&\leq& {\frac{n}{\nu_b\varrho}}\sum_{j\neq b}\abs{\e_j^T\matV\matV^T\e_b}^2\left(\frac{1}{c_0\text{ log}(m+n)}\cdot \frac{n}{\nu_j\varrho}\sum_{i}\matZ_{ij}^2\right)\\
%
%
&\leq& \left(\frac{\muTNorm{\matZ}^2}{c_0\text{ log}(m+n)}\right){\frac{n}{\nu_b\varrho}}\sum_{j\neq b}\abs{\e_j^T\matV\matV^T\e_b}^2\\
&\leq& \frac{\muTNorm{\matZ}^2}{c_0\text{ log}(m+n)},
\end{eqnarray*}
where the last inequality follows from, 
$
\sum_{j\neq b}\abs{\e_j^T\matV\matV^T\e_b}^2 
\leq \TNormS{\matV\matV^T\e_b} 
\leq \frac{\nu_b\varrho}{n}
$.\\
Combining the two summations,
\begin{eqnarray*}
\TNorm{\sum_{i,j}\mathbb{E}\left[\s_{ij}^T\s_{ij}\right]} \leq \frac{5}{c_0\text{ log}(m+n)}\muTNorm{\matZ}^2
\end{eqnarray*}
We can bound $\TNorm{\mathbb{E}\left[\sum_{i,j}\s_{ij}\s_{ij}^T\right]}$ in a similar way.\\
We apply Matrix Bernstein inequality in Lemma \ref{thm:matrix_bernstein}, with 
$$
\gamma = \frac{2}{c_0\text{ log}(m+n)}\muNorm{\matZ}, 
\quad \sigma^2 = \frac{5}{c_0\text{ log}(m+n)}\muTNorm{\matZ}^2,
$$
to obtain
$$
\TNorm{\sum_{i,j}\s_{ij}} \leq \sqrt{\frac{20c}{c_0}}\muTNorm{\matZ} + \frac{2c}{c_0}\muNorm{\matZ}.
$$
We set 
$c_0 \geq {80c}$
to derive
$$
\TNorm{\sqrt{\frac{n}{\nu_b\varrho}}\matX_{*, b}} = \TNorm{\sum_{i,j}\s_{ij}} \leq \frac{1}{2}\left(\muTNorm{\matZ} + \muNorm{\matZ}\right).
$$
Similarly, we can bound $\TNorm{\sqrt{\frac{m}{\mu_a\varrho}}\matX_{a, * }}$ by the same quantity. We take a union bound over all rows $a$ and all columns $b$ (i.e., total $(m+n)$ events) to obtain, for any $c>2$,
\begin{eqnarray*}
\muTNorm{(\mathcal{P}_T\mathcal{R}_{\Omega} - \mathcal{P}_T)(\matZ)} \leq \frac{1}{2}\left(\muTNorm{\matZ} + \muNorm{\matZ}\right)
\end{eqnarray*}
holding with probability at least $1 - (m+n)^{2-c}$.


%% file: lemma_12_our_modified.tex
%
%
Let, $\matX = (\mathcal{P}_T\mathcal{R}_{\Omega} - \mathcal{P}_T)\matZ = \sum_{i,j}\left(\frac{\delta_{ij}}{q_{ij}}-1\right)\matZ_{ij}\left(\mathcal{P}_T(\e_i\e_j^T)\right)$.
We write rescaled $(a,b)$-th element of $\matX$ as
$$
[\matX]_{ab}\sqrt{\frac{m}{\mu_a\varrho}}\sqrt{\frac{n}{\nu_b\varrho}} = \sum_{i,j}\left(\frac{\delta_{ij}}{q_{ij}}-1\right)\matZ_{ij}\left(\mathcal{P}_T(\e_i\e_j^T)\right)_{ab}\sqrt{\frac{m}{\mu_a\varrho}}\sqrt{\frac{n}{\nu_b\varrho}} = \sum_{i,j}s_{ij}
$$
This is a sum of independent, zero-mean random variables. we seek to bound $\abs{s_{ij}}$ and $\abs{\sum_{i,j}\mathbb{E}\left[s_{ij}^2\right]}$.
%
First, we need to bound $\abs{\left<\e_a\e_b^T,\mathcal{P}_T(\e_i\e_j^T)\right>}$.
%
%
%
\begin{eqnarray}
\nonumber
&&\abs{\left<\e_a\e_b^T,\mathcal{P}_T(\e_i\e_j^T)\right>}\\
\nonumber
&=& \abs{\e_a^T\matU\matU^T(\e_i\e_j^T)\e_b + \e_a^T(\e_i\e_j^T)\matV\matV^T\e_b - \e_a^T\matU\matU^T(\e_i\e_j^T)\matV\matV^T\e_b}\\
%
%
&=&
\begin{cases}
\FNormS{\mathcal{P}_T(\e_a\e_b^T)} = \frac{\mu_a\varrho}{m}+\frac{\nu_b\varrho}{n} - \frac{\mu_a\varrho}{m}\cdot \frac{\nu_b\varrho}{n}& i=a, j=b,\\
%
\abs{\e_a^T(\matI - \matU\matU^T)\e_a\e_j^T\matV\matV^T\e_b} \leq \abs{\e_j^T\matV\matV^T\e_b} & i=a, j\neq b,\\
\abs{\e_a^T\matU\matU^T\e_i\e_b^T(\matI - \matV\matV^T)\e_b} \leq \abs{\e_a^T\matU\matU^T\e_i} & i\neq a, j=b,\\
\abs{\e_a^T\matU\matU^T\e_i\e_j^T\matV\matV^T\e_b} \leq \abs{\e_a^T\matU\matU^T\e_i}\abs{\e_j^T\matV\matV^T\e_b} & i\neq a, j \neq b\\
\end{cases}
\end{eqnarray}
where we use $\TNorm{\matI - \matU\matU^T}\leq 1$ and $\TNorm{\matI - \matV\matV^T}\leq 1$.
\\
%
%
%
%
%
%
Note that, $s_{ij} = 0$ when $q_{ij}=1$. Otherwise, for $q_{ij} \neq 1$,
%
$$
\abs{s_{ij}} \leq \frac{1}{q_{ij}}\abs{\matZ_{ij}}\abs{\left< \e_a\e_b^T,\mathcal{P}_T(\e_i\e_j^T)\right>}\sqrt{\frac{m}{\mu_a\varrho}}\sqrt{\frac{n}{\nu_b\varrho}}
$$
We consider four cases.
\\
For $i=a, j=b$, using 
$q_{ab} \geq c_0\text{log}(m+n)\left(\frac{\mu_a\varrho}{m} + \frac{\nu_b\varrho}{n}-\frac{\mu_a\varrho}{m} \cdot \frac{\nu_b\varrho}{n}\right)$
\begin{eqnarray*}
\abs{s_{ij}} 
&\leq& \frac{1}{q_{ab}}\abs{\matZ_{ab}}\FNormS{\mathcal{P}_T(\e_a\e_b^T)}\sqrt{\frac{m}{\mu_a\varrho}}\sqrt{\frac{n}{\nu_b\varrho}}\\
&\leq& \frac{\abs{\matZ_{ab}}}{c_0\text{ log}(m+n)}\sqrt{\frac{m}{\mu_a\varrho}}\sqrt{\frac{n}{\nu_b\varrho}} 
\leq \frac{\muNorm{\matZ}}{c_0\text{ log}(m+n)}
\end{eqnarray*}
For $i=a, j\neq b$, using 
$q_{aj} \geq c_0\text{ log}(m+n)\left(\frac{\mu_a\varrho}{m} + \frac{\nu_j\varrho}{n} - \frac{\mu_a\varrho}{m} \cdot \frac{\nu_j\varrho}{n}\right)\geq c_0\text{ log}(m+n)\frac{\nu_j\varrho}{n}
$,
\begin{eqnarray*}
\abs{s_{ij}} \leq \frac{\abs{\matZ_{aj}}}{q_{aj}}\abs{\e_j^T\matV\matV^T\e_b}\sqrt{\frac{m}{\mu_a\varrho}}\sqrt{\frac{n}{\nu_b\varrho}}
%
%
\leq \frac{\abs{\matZ_{aj}}}{c_0\text{ log}(m+n)}\sqrt{\frac{n}{\nu_j\varrho}}\sqrt{\frac{m}{\mu_a\varrho}}
\leq \frac{\muNorm{\matZ}}{c_0\text{log}(m+n)}
\end{eqnarray*}
Similarly, for $i\neq a, j=b$, using  $q_{ib} \geq  c_0\text{log}(m+n)\frac{\mu_i\varrho}{m}$
$$\abs{s_{ij}} \leq \frac{\muNorm{\matZ}}{c_0\text{ log}(m+n)}.$$
For $i\neq a, j \neq b$, using $q_{ij} \geq c_0\text{ log}(m+n)\sqrt{\frac{\mu_i\varrho}{m}\cdot \frac{\nu_j\varrho}{n}}$
%
%
%
\begin{eqnarray*}
\abs{s_{ij}} &\leq& \frac{1}{q_{ij}}\abs{\matZ_{ij}} \abs{\e_a^T\matU\matU^T\e_i}\abs{\e_j^T\matV\matV^T\e_b}
\sqrt{\frac{m}{\mu_a\varrho}}\sqrt{\frac{n}{\nu_b\varrho}}\\
&\leq& \frac{1}{q_{ij}}\abs{\matZ_{ij}} \sqrt{\frac{\mu_i\varrho}{m}} \sqrt{\frac{\mu_a\varrho}{m}}\cdot \sqrt{\frac{\nu_b\varrho}{n}}\sqrt{\frac{\nu_j\varrho}{n}}\cdot\sqrt{\frac{m}{\mu_a\varrho}}\sqrt{\frac{n}{\nu_b\varrho}}\\
%
%
%
&\leq& \frac{1}{c_0\text{ log}(m+n)}\abs{\matZ_{ij}}\sqrt{\frac{m}{\mu_i\varrho}}\sqrt{\frac{n}{\nu_j\varrho}}
\leq \frac{1}{c_0\text{ log}(m+n)}\muNorm{\matZ}.
\end{eqnarray*}
Above we use $\sqrt{\frac{\mu_i\varrho}{m}}\leq 1$, $\sqrt{\frac{\nu_j\varrho}{n}}\leq 1$, for all $i$, $j$.
We conclude, for all $(i,j)$,
$$
\abs{s_{ij}} \leq \frac{1}{c_0\text{ log}(m+n)}\muNorm{\matZ}.
$$
On the other hand,
\begin{eqnarray*}
\abs{\sum_{i,j}\mathbb{E}\left[s^2_{ij}\right]} 
&=& \sum_{i,j}\mathbb{E}\left[\left(\frac{\delta_{ij}}{q_{ij}}-1\right)^2\right]\matZ_{ij}^2{\left< \e_a\e_b^T,\mathcal{P}_T(\e_i\e_j^T)\right>}^2\frac{m}{\mu_a\varrho}\cdot \frac{n}{\nu_b\varrho}\\
&=& \sum_{i,j}\frac{1-q_{ij}}{q_{ij}}\matZ_{ij}^2{\left< \e_a\e_b^T,\mathcal{P}_T(\e_i\e_j^T)\right>}^2\frac{m}{\mu_a\varrho}\cdot \frac{n}{\nu_b\varrho}\\
%
%
&=& \sum_{i=a,j=b}+\sum_{i=a,j\neq b}+\sum_{i\neq a,j=b}+\sum_{i\neq a,j\neq b}
\end{eqnarray*}
The above quantity is zero for $q_{ij}=1$. We bound the above considering four cases for $q_{ij}\neq 1$.
\\
For $i=a, j=b$, using 
$q_{ab} \geq c_0\text{ log}(m+n)\left(\frac{\mu_a\varrho}{m}+\frac{\nu_b\varrho}{n} - \frac{\mu_a\varrho}{m}\cdot\frac{\nu_b\varrho}{n}\right) 
$,
\begin{eqnarray*}
\sum_{i=a,j=b} 
%
%
\leq \frac{\matZ_{ab}^2}{q_{ab}}\left(\frac{\mu_a\varrho}{m}+\frac{\nu_b\varrho}{n}-\frac{\mu_a\varrho}{m}\cdot\frac{\nu_b\varrho}{n}\right)^2\frac{m}{\mu_a\varrho}\cdot \frac{n}{\nu_b\varrho}
%
%
\leq \frac{\muNorm{\matZ}^2}{c_0\text{ log}(m+n)}
\end{eqnarray*}
Above we use ${\left(\frac{\mu_i\varrho}{m}+\frac{\nu_j\varrho}{n} - \frac{\mu_i\varrho}{m}\cdot\frac{\nu_j\varrho}{n}\right)} \leq 1$, for all $i$ and $j$.\\
\\
For $i=a, j\neq b$, using 
$q_{aj} \geq c_0\text{ log}(m+n)\frac{\nu_j\varrho}{n}$,
%
%
\begin{eqnarray*}
\sum_{i=a,j\neq b}&\leq& \sum_{j\neq b}\frac{1}{q_{aj}}\matZ_{aj}^2 \abs{\e_j^T\matV\matV^T\e_b}^2\frac{m}{\mu_a\varrho}\frac{n}{\nu_b\varrho}\\
&\leq& \frac{1}{ c_0\text{ log}(m+n)}\sum_{j\neq b}\matZ_{aj}^2\left(\frac{n}{\nu_j\varrho}\frac{m}{\mu_a\varrho}\right)\abs{\e_j^T\matV\matV^T\e_b}^2\frac{n}{\nu_b\varrho}\\
%
&\leq& \frac{1}{ c_0\text{ log}(m+n)}\muNorm{\matZ}^2.
\end{eqnarray*}
Above we use,
$$
\sum_{j\neq b}\abs{\e_j^T\matV\matV^T\e_b}^2 \leq \TNormS{\matV\matV^T\e_b} 
\leq \frac{\nu_b\varrho}{n}.
$$
Similarly, we can derive identical bound for $\sum_{i\neq a, j=b}$.\\
%
%
We use 
$
q_{ij} 
\geq c_0\text{ log}(m+n)\sqrt{\frac{\mu_i\varrho}{m}\cdot \frac{\nu_j\varrho}{n}}
\geq c_0\text{ log}(m+n){\frac{\mu_i\varrho}{m}}\cdot {\frac{\nu_j\varrho}{n}}
$
to bound
%
%
%
\begin{eqnarray*}
\sum_{i\neq a,j\neq b}&\leq& \sum_{i\neq a, j\neq b}\frac{1}{q_{ij}}\matZ_{ij}^2 \abs{\e_a^T\matU\matU^T\e_i}^2\abs{\e_j^T\matV\matV^T\e_b}^2\frac{m}{\mu_a\varrho}\frac{n}{\nu_b\varrho}\\
%
%
&\leq& \frac{\muNorm{\matZ}^2}{c_0\text{ log}(m+n)}\sum_{i\neq a, j\neq b} \abs{\e_a^T\matU\matU^T\e_i}^2\abs{\e_j^T\matV\matV^T\e_b}^2\frac{m}{\mu_a\varrho}\frac{n}{\nu_b\varrho}\\
&=& \frac{\muNorm{\matZ}^2}{c_0\text{ log}(m+n)}\sum_{i\neq a} \abs{\e_a^T\matU\matU^T\e_i}^2\frac{m}{\mu_a\varrho}\sum_{j\neq b}\abs{\e_j^T\matV\matV^T\e_b}^2\frac{n}{\nu_b\varrho}\\
&\leq& \frac{\muNorm{\matZ}^2}{c_0\text{ log}(m+n)}
\end{eqnarray*}
Combining the summations, we derive
%
$$
\abs{\sum_{i,j}\mathbb{E}\left[s^2_{ij}\right]} \leq \frac{4\muNorm{\matZ}^2}{c_0\text{ log}(m+n)}.
$$
%
We now apply Bernstein inequality in Lemma \ref{thm:matrix_bernstein}
%
to obtain, for any $c>3$, $c_0 \geq 68c$
\begin{eqnarray*}
\muNorm{(\mathcal{P}_T\mathcal{R}_{\Omega} - \mathcal{P}_T)(\matZ)} \leq \frac{1}{2}\muNorm{\matZ}
\end{eqnarray*}
We take union bound over all $(a,b)$ (i.e., total $mn \leq (m+n)^2$ events) to conclude that the above result holds with probability at least
$$
1 - (m+n)^{(3-c)}.
$$

%% file: theorem_2_our_modified.tex

\subsection{Proof of Lemma \ref{lem:row_sample}}

Let $\delta_i = \mathbb{I}(i \in \Gamma)$, where $\mathbb{I}(\cdot)$ is the indicator function. We can write,
$$
\matU^T\mathcal{S}_\Gamma(\matU) - \matI_{\varrho} 
=
\matU^T\mathcal{S}_\Gamma(\matU) - \matU^T\matU
=
\sum_{i=1}^{m}\left(\frac{1}{p}\delta_i - 1\right)\matU^T\e_i\e_i^T\matU
= 
\sum_{i=1}^{m}\matS_i,
$$
where $\matS_i$'s are independent of each other.
Clearly, $\mathbb{E}[\matS_i] = 0_{m \times n}$.\\
Note that,
$\TsNorm{\matS_i}^2 \leq \frac{1}{p}\TsNorm{\matU^T\e_i}^2 \leq \frac{\mu_0\varrho}{pm}$.
Also,
\begin{eqnarray*}
\TNorm{\mathbb{E}\left[\sum_{i=1}^{m}\matS_i\matS_i^T\right]} = \TNorm{\mathbb{E}\left[\sum_{i=1}^{m}\matS_i^T\matS_i\right]}
&=&
\frac{1-p}{p} \TNorm{\sum_{i=1}^{m}\matU^T\e_i\e_i^T\matU\matU^T\e_i\e_i^T\matU}\\
&=&
\frac{1-p}{p} \TNorm{\matU^T\left(\sum_{i=1}^{m}\e_i\e_i^T\matU\matU^T\e_i\e_i^T\right)\matU}\\
&\leq& 
\frac{1}{p} \TNorm{\sum_{i=1}^{m}\e_i\e_i^T\TsNorm{\matU^T\e_i}^2}\\
&\leq& 
\frac{1}{p} \max_i\TsNorm{\matU^T\e_i}^2 \leq \frac{\mu_0\varrho}{pm}
\end{eqnarray*}
We apply the matrix Bernstein inequality in Lemma \ref{thm:matrix_bernstein} to derive the result. 

%% file: conclusion.tex
\\

\textbf{Acknowledgment:} 
I thank Prof Petros Drineas and Prof Malik Magdon-Ismail for helpful discussions on this topic. Prof Drineas encouraged me to work on this problem.

%% file: Elementwise_relaxed_Leverage_NNM.bbl
\begin{thebibliography}{10}

\bibitem{CLMW11}
E.~J. Candes, X.~Li, Y.~Ma, and J.~Wright.
\newblock {Robust principal component analysis?}
\newblock In {\em Journal of the ACM}, page 58(3):11, 2011.

\bibitem{CR09}
E.~J. Candes and B.~Recht.
\newblock {Exact matrix completion via convex optimization}.
\newblock In {\em Foundations of Computational Mathematics}, pages 717–--772,
  9(6), 2009.

\bibitem{CT10}
E.~J. Candes and T.~Tao.
\newblock {The power of convex relaxation: Near-optimal matrix completion}.
\newblock In {\em IEEE Transactions on Information Theory}, pages 2053–--2080,
  56(5), 2010.

\bibitem{CH86}
S.~Chatterjee and A.~Hadi.
\newblock {Influential observations, high leverage points, and outliers in
  linear regression}.
\newblock {\em Statistical Science}, pages 1(3):379--393, 1986.

\bibitem{BCSW14}
Y~Chen, S~Bhojanapalli, S~Sanghavi, and R~Ward.
\newblock {Completing Any Low-rank Matrix, Provably}.
\newblock {\em \url{http://arxiv.org/abs/1306.2979}}, 2014.

\bibitem{DMM08}
P.~Drineas, M.W. Mahoney, and S.~Muthukrishnan.
\newblock {Relative-error CUR matrix decompositions}.
\newblock In {\em SIAM Journal on Matrix Analysis and Applications}, pages
  844--881, 30(2), 2008.

\bibitem{Fazel02}
M.~Fazel.
\newblock Matrix rank minimization with applications.
\newblock In {\em PhD thesis, Stanford University}, 2002.

\bibitem{GM04}
E.~Gabrilovich and S.~Markovitch.
\newblock {Text categorization with many redundant features: using aggressive
  feature selection to make SVMs competitive with C4.5}.
\newblock In {\em Proceedings of International Conference on Machine Learning},
  2004.

\bibitem{Gross11}
D.~Gross.
\newblock {Recovering low-rank matrices from few coefficients in any basis}.
\newblock In {\em IEEE Transactions on Information Theory}, pages 1548–--1566,
  57(3), 2011.

\bibitem{KS13}
A.~Krishnamurthy and A.~Singh.
\newblock Low-rank matrix and tensor completion via adaptive sampling.
\newblock In {\em Advances in Neural Information Processing Systems 26}, pages
  836--844. 2013.

\bibitem{MD09}
M.W. Mahoney and P.~Drineas.
\newblock {CUR matrix decompositions for improved data analysis}.
\newblock In {\em Proceedings of the National Academy of Sciences}, pages
  697--702, 106 (3), 2009.

\bibitem{Natarajan95}
B.K. Natarajan.
\newblock {Sparse approximate solutions to linear systems}.
\newblock pages 227–--234, 24, 1995.

\bibitem{Recht09}
B.~Recht.
\newblock {A simpler approach to matrix completion}.
\newblock In {\em The Journal of Machine Learning Research}, pages 3413--3430,
  12, 2011.

\bibitem{Tropp12}
J.~Tropp.
\newblock {User-friendly tail bounds for sums of random matrices}.
\newblock In {\em Foundations of Computational Mathematics}, pages
  12(4):389--434, 2012.

\end{thebibliography}
